\documentclass[twocolumn,10pt,amsmath,amssymb,pra]{revtex4-1}
\usepackage{graphicx}
\usepackage{longtable}
\usepackage{color}
\usepackage{dcolumn}
\usepackage{multirow}
\usepackage{epstopdf}

\newcolumntype{d}[1]{D{.}{.}{#1}}

\newcommand{\idest}[1]
{
  \textit{i.e.~}#1
}

\newcommand{\sixj}[6]
{
  \left\{ \begin{array}{ccc} #1 & #2 & #3 \\
    #4 & #5 & #6 \end{array} \right\}
}
\newcommand{\thrj}[6]
{
  \left( \begin{array}{ccc} #1 & #2 & #3 \\
    #4 & #5 & #6 \end{array} \right)
}
\newcommand{\dip}[2]
{
  \left\langle #1 \left\Vert \mathbf{d} 
  \right\Vert #2 \right\rangle
}
\newcommand{\dipsq}[2]
{
  \left| \dip{#1}{#2} \right|^2
}

\begin{document}

\title{Anisotropic optical trapping as a manifestation of the complex electronic structure of ultracold lanthanide atoms: the example of holmium}

\author{Hui Li$^{1}$, Jean-Fran{\c c}ois Wyart$^{1,2}$, Olivier Dulieu$^{1}$ and Maxence Lepers$^{1}$}
\email{maxence.lepers@u-psud.fr}
\address{${}^{1}$Laboratoire Aim{\'e} Cotton, CNRS, Universit{\'e} Paris-Sud, ENS Cachan, Universit{\'e} Paris-Saclay, 91405 Orsay, France}
\address{${}^{2}$LERMA, Observatoire de Paris-Meudon, PSL Research University, Sorbonne Universit{\'e}s, UPMC Univ.~Paris 6, CNRS UMR8112, 92195 Meudon, France}

\begin{abstract}
The efficiency of optical trapping is determined by the atomic dynamic dipole polarizability, whose real and imaginary parts are associated with the potential energy and photon-scattering rate respectively. In this article we develop a formalism to calculate analytically the real and imaginary parts of the scalar, vector and tensor polarizabilities of lanthanide atoms. We assume that the sum-over-state formula only comprises transitions involving electrons in the valence orbitals like $6s$, $5d$, $6p$ or $7s$, while transitions involving $4f$ core electrons are neglected. Applying this formalism to the ground level of configuration $4f^q6s^2$, we restrict the sum to transitions implying the $4f^q6s6p$ configuration, which yields polarizabilities depending on two parameters: an effective transition energy and an effective transition dipole moment. Then, by introducing configuration-interaction mixing between $4f^q6s6p$ and other configurations, we demonstrate that the imaginary part of the scalar, vector and tensor polarizabilities is very sensitive to configuration-interaction coefficients, whereas the real part is not. The magnitude and anisotropy of the photon-scattering rate is thus strongly related to the details of the atomic electronic structure. Those analytical results agree with our detailed electronic-structure calculations of energy levels, Land\'e $g$-factors, transition probabilities, polarizabilities and van der Waals $C_6$ coefficients, previously performed on erbium and dysprosium, and presently performed on holmium. Our results show that, although the density of states decreases with increasing  $q$, the configuration interaction between $4f^q6s6p$, $4f^{q-1}5d6s^2$ and $4f^{q-1}5d^26s$ is surprisingly stronger in erbium ($q=12$), than in holmium ($q=11$), itself stronger than in dysprosium ($q=10$).
\end{abstract}


\maketitle

\section{Introduction}

The physics of ultracold gases has evolved rapidly and is poised to enter a new promising regime, where complex atomic and molecular species can be cooled and studied extensively. Lanthanide atoms, with a strong magnetic moment and a large orbital angular momentum, are extreme examples of such complex species. In fact, the interest for ultracold lanthanide atoms is motivated by several topics in current research, including ultracold collisions and quantum chaos \cite{frisch2014, maier2015, tang2015b}, dipolar quantum gases with large magnetic moment and strong dipole-dipole interaction \cite{lu2012, nessi2014, frisch2015, yao2015, baier2016}, many-body quantum systems \cite{maier2015b, burdick2015}, exotic quantum phases \cite{fregoso2009, aikawa2014b, kadau2016} like stable quantum droplets \cite{ferrier2016, xi2016, macia2016},  synthetic gauge field \cite{cui2013, burdick2016}, and optical clocks \cite{kozlov2013, vishnyakova2014, sukachev2016}. Recent progress in laser cooling and magneto-optical trapping of high-atomic-number (high-$Z$) lanthanides \cite{hancox2004, hemmerling2014}, including dysprosium (Dy) \cite{leefer2010, lu2010, maier2014, dreon2016}, erbium (Er) \cite{ban2005, mcclelland2006, frisch2012}, holmium (Ho) \cite{miao2014} and thulium (Tm) \cite{sukachev2010} is paving the way towards these investigations. In  addition, both Bose-Einstein condensates and quantum-degenerate Fermi gases have been produced in isotopes of Dy \cite{lu2012, lu2011b,  tang2015} and Er \cite{aikawa2012, aikawa2014}. 

The ground level of holmium is characterized by the electronic configuration [Xe]$4f^{11}6s^2$ and an electronic angular momentum $J=15/2$. Due to the nuclear spin $I=7/2$ of its only stable (bosonic) isotope $^{165}$Ho, holmium is the atom possessing the largest number of hyperfine sublevels in the electronic ground level, namely $(2J+1) \times (2I+1) = 128$. This rich structure is likely to be exploited in quantum information \cite{saffman2008, hostetter2015}. Like other lanthanides, the complex electronic structure of holmium induces a large magnetic dipole moment (9\;$\mu_B$) that makes it an interesting candidate to investigate the anisotropic interactions between atoms \cite{newman2011, kao2016}. Recently the holmium single magnetic atom and holmium molecular nanomagnet was also presented as a competing candidate for the realization of quantum bits \cite{miyamachi2013, shiddiq2016}.

Many of the applications listed above involve optically trapped ultracold atoms. The trapping efficiency is determined by the interaction between the atoms and the electromagnetic field \cite{manakov1986, grimm2000}. The microscopic property characterizing the atomic response is the (complex) dynamic dipole polarizability (DDP). On the one hand, the field induces a potential energy,\idest{an}ac-Stark shift, on the atoms, which is proportional to the real part of the DDP. On the other hand, the field also induces photon-scattering, whose rate is proportional to the imaginary part of the DDP. In ultracold experiments, it is necessary to characterize the photon-scattering rate, as it provokes heating of the sample, and trap losses \cite{grimm2000}. Beyond trapping itself, the real part of the vector and tensor DDPs are also necessary to determine the Raman-coupling strengths between different Zeeman sublevels, which was proposed for the implementation of synthetic gauge fields \cite{cui2013, burdick2016}. In our previous works on Er \cite{lepers2014} and Dy \cite{li2017}, we have shown that, far from resonant frequencies, the ac-Stark shift only weakly depends on the field polarization and atomic Zeeman sublevel, despite the absence of spherical symmetry in the $4f$-electron wave functions. We have unveiled the inverse situation for photon-scattering, as the imaginary part of the vector and tensor DDPs represent significant fractions of the scalar one. This opens the possibility to control the trap heating and losses with an appropriate field polarization. However, the vector-to-scalar and tensor-to-scalar ratio vary strongly from Dy to Er, which is still unexplained.

Understanding the origin of that difference is a major motivation of the present work. Moreover, ultracold experiments may require to characterize the optical trapping of atomic excited levels with energies up to 25000 cm$^{-1}$ above the ground level. Calculating the DDP of such levels with the sum-over-state formula requires to model highly-excited levels, roughly up to 60000 cm$^{-1}$ above the ground level, which is a hard task for the most complex spectra of lanthanide atoms. Therefore, in this article, we present a simplified model of the DDP based on the sum-over-state formula, where we suppose that the only contributions come from transitions  involving valence electrons like $6s$, $6p$, $5d$ or $7s$, and where we ignore transitions involving $4f$ core electrons. Assuming that all the levels of a given configuration have similar energies, we obtain analytical expressions of the DDPs of an arbitrary level, depending on a restricted number of effective parameters. Focusing on the ground level, of configuration [Xe]$\,4f^q6s^2$ ($q=10$, 11 and 12 for Dy, Ho and Er, respectively), we only take into account the excitation from the $6s$ to the $6p$ orbital, but not the excitation from the $4f$ to the $5d$ orbital. We demonstrate that the real part of the DDP is not influenced by the configuration interaction (CI) between [Xe]$\,4f^q6s6p$ and other configurations like [Xe]$4f^{q-1} 5d 6s^2$ and [Xe]$4f^{q-1} 5d^2 6s$. Our model also shows that the real part of the vector and tensor ground-level DDPs vanish. By contrast, the imaginary part of the DDPs is very sensitive to CI, and in particular to the weight of the [Xe]$4f^q6s6p$ configuration in excited levels. We demonstrate that a strong CI mixing tends to increase the vector and tensor DDPs with respect to the scalar one. Surprisingly, CI mixing turns out to be larger for Er than for Ho, and for Ho than for Dy, although the energy spectrum of Dy is the densest one.

In order to check the validity of those conclusions, we perform a full numerical modeling of holmium spectrum, including energy levels, transition probabilities, polarizabilities and van der Waals $C_6$ coefficients, complementing our previous studies on erbium \cite{lepers2014} and dysprosium \cite{li2017}. The DDPs and $C_6$ coefficients are calculated using the sum formula involving transition energies and transition dipole moments extracted from our computed transition probabilities. Following our previous work \cite{wyart2011, lepers2014, lepers2016, li2017}, those quantities are calculated using a combination of \textit{ab initio} and least-square fitting procedures provided by the Cowan suite of codes \cite{cowan1981} and extended in our group. Therefore we provide a theoretical interpretation of Ho even-parity levels, which especially results in the prediction of the widely unmeasured Land\'e $g$-factors. Because the spectrum of high-$Z$ lanthanide atoms in the ground level is composed of a few strong transitions emerging from a forest of weak ones, the sum-over-state formula is appropriate to calculate DDPs and $C_6$ coefficients. It offers the possibility to precisely calculate, with a single set of spectroscopic data, the real and imaginary parts of the scalar, vector and tensor DDPs, in a wide range of wavelengths, especially at 1064 nm, widely used experimentally for trapping purposes.

This article is outlined as follows. We develop our simplified model for the DDP in section \ref{sec:simple}: we first recall useful formulas and especially the relationships between scalar, vector and tensor DDPs and tensor operators (see subsection \ref{sub:tens-op}). Then we calculate the contribution from the levels of a single configuration (see subsection \ref{sub:1array}) to the real and imaginary parts of the DDPs, while the two next subsections are devoted to the influence of CI mixing in the DDPs of the ground level of lanthanide atoms. The second part of the paper (section \ref{sec:ho-spec}) deals with the full numerical modeling of holmium spectrum -- energy levels, transition probabilities, polarizabilities and van der Waals $C_6$ coefficients (see subsections \ref{sub:ener-lev}--\ref{sub:vdw} respectively). Section \ref{sec:conclu} contains concluding remarks.

\section{Dynamic dipole polarizability: a simplified model}
\label{sec:simple}

\subsection{Polarizability and tensor operators}
\label{sub:tens-op}

For non-spherically-symmetric atoms like lanthanides, the ac-Stark shift is a linear combination of three terms, depending on the scalar, vector and tensor polarizabilities, taken at the angular frequency $\omega$ of the oscillating electric field (hereafter denoted {}``frequency''). The magnitude of each term  is a function of the atomic Zeeman sublevel $M$ and of the electric-field polarization \cite{manakov1986}. The scalar $\alpha_\mathrm{scal}(\omega)$, vector $\alpha_\mathrm{vect}(\omega)$ and tensor polarizabilities $\alpha_\mathrm{tens}(\omega)$ can be associated with the coupled polarizabilities $\alpha_{k}(\omega)$, where $k=0$, 1 and 2 respectively, is the rank of the corresponding irreducible tensor \cite{manakov1986, beloy2009}. Namely 
\begin{eqnarray}
  \alpha_\mathrm{scal}(\omega) & = & -\frac{\alpha_{0}(\omega)}
    {\sqrt{3(2J+1)}}
  \label{eq:alpha-scal} \\
  \alpha_\mathrm{vect}(\omega) & = & \alpha_{1}(\omega)
    \sqrt{\frac{2J}{(J+1)(2J+1)}}
  \label{eq:alpha-vect} \\
  \alpha_\mathrm{tens}(\omega) & = & \alpha_{2}(\omega)
    \sqrt{\frac{2J(2J-1)}{3(J+1)(2J+1)(2J+3)}} \,.
  \label{eq:alpha-tens}
\end{eqnarray}
For an atomic level $|\beta J\rangle$, where $J$ is the electronic-angular-momentum quantum number and $\beta$ stands for all the other quantum numbers, the general expression for $\alpha_{k}$ is
\begin{align}
  \alpha_{k}(\omega) = & \sqrt{2k+1} 
   \sum_{\beta''J''} (-1)^{J+J''}
  \nonumber \\
   \times & \sixj{1}{1}{k}{J}{J}{J''} \dipsq{\beta''J''}{\beta J}
  \nonumber \\
   \times & \left( \frac{ (-1)^k}
    {E_{\beta''J''}-E_{\beta J} 
     -i\frac{\hbar\gamma_{\beta''J''}}{2}-\hbar\omega} \right.
  \nonumber \\
   & \left. + \frac{1} {E_{\beta''J''}-E_{\beta J} 
     -i\frac{\hbar\gamma_{\beta''J''}}{2}+\hbar\omega} \right)
  \label{eq:alpha-k}
\end{align}
where $E_{\beta J}$ ($E_{\beta''J''}$) are the energies of the levels $|\beta J\rangle$ ($|\beta''J''\rangle$), $\dip{\beta''J''}{\beta J}$ is the reduced transition dipole moment between these two levels, $\gamma_{\beta''J''}$ is the natural linewidth of the intermediate level $|\beta'' J''\rangle$, and the quantity between curly brackets is a Wigner 6-j symbol \cite{varshalovich1988}.

We consider frequencies far from any atomic resonances, \textit{i.e.}~$E_{\beta''J''} - E_{\beta J} \pm \hbar\omega \gg \hbar\gamma_{\beta''J''}/2$, which is relevant for trapping purposes, and which greatly simplfies Eq.~(\ref{eq:alpha-k}). We separate the real $\Re[\alpha_k(\omega)]$ and imaginary parts $\Im[\alpha_k(\omega)]$,
\begin{widetext}
\begin{align}
  \Re[\alpha_{k}(\omega)] & = 2 \sqrt{2k+1} 
   \sum_{\beta''J''} (-1)^{J+J''}
   \sixj{1}{1}{k}{J}{J}{J''} \dipsq{\beta''J''}{\beta J}
   \frac{(E_{\beta''J''}-E_{\beta J}) 
    \delta_{(-1)^k,1} - \hbar\omega \delta_{(-1)^k,-1}}
    {(E_{\beta''J''}-E_{\beta J})^2 - \hbar^2\omega^2}
  \label{eq:alpha-k-re} \\
  \Im[\alpha_{k}(\omega)] & = \sqrt{2k+1} 
   \sum_{\beta''J''} (-1)^{J+J''} \sixj{1}{1}{k}{J}{J}{J''} 
   \hbar\gamma_{\beta''J''} \dipsq{\beta''J''}{\beta J}
  \nonumber \\
   & \times \frac{[ (E_{\beta''J''}-E_{\beta J})^2 
    + \hbar^2 \omega^2 ] \delta_{(-1)^k,1}
    - 2\hbar\omega(E_{\beta''J''}-E_{\beta J}) \delta_{(-1)^k,-1}}
    {\left[(E_{\beta''J''}-E_{\beta J})^2 - 
     \hbar^2\omega^2\right]^2},
  \label{eq:alpha-k-im}
\end{align}
where we used $A + (-1)^kB = (A+B)\delta_{(-1)^k,1} + (A-B) \delta_{(-1)^k,-1}$. Plugging Eqs.~\eqref{eq:alpha-k-re} and \eqref{eq:alpha-k-im} into Eqs.~(\ref{eq:alpha-scal})--(\ref{eq:alpha-tens}), and introducing the explicit expressions of 6-j symbols (see Ref.~\cite{varshalovich1988}, p.~302), we get to the real and imaginary parts of the scalar, vector and tensor contributions,
\begin{align}
  \Re[\alpha_\mathrm{scal}(\omega)] & = \frac{2}{3(2J+1)}
    \sum_{\beta''J''} \frac{ (E_{\beta''J''}-E_{\beta J})
      \dipsq{\beta''J''}{\beta J}}
    {(E_{\beta''J''}-E_{\beta J})^2-\hbar^2\omega^2}
  \label{eq:alpha-scal-re} \\
  \Im[\alpha_\mathrm{scal}(\omega)] & = \frac{1}{3(2J+1)}
    \sum_{\beta''J''} \frac{ \hbar\gamma_{\beta''J''}
      \left[(E_{\beta''J''}-E_{\beta J})^2+\hbar^2\omega^2\right]
      \dipsq{\beta''J''}{\beta J}}
    {\left[(E_{\beta''J''}-E_{\beta J})^2 
    -\hbar^2\omega^2\right]^2}
  \label{eq:alpha-scal-im} \\
  \Re[\alpha_\mathrm{vect}(\omega)] & = \sum_{\beta''J''}
   \frac{J''(J''+1)-J(J+1)-2}{(J+1)(2J+1)} \times
   \frac{\hbar\omega \dipsq{\beta''J''}{\beta J}}
    {(E_{\beta''J''}-E_{\beta J})^2-\hbar^2\omega^2}  
  \label{eq:alpha-vect-re} \\
  \Im[\alpha_\mathrm{vect}(\omega)] & = \sum_{\beta''J''}
   \frac{J''(J''+1)-J(J+1)-2}{(J+1)(2J+1)} \times
   \frac{\hbar^2 \omega \gamma_{\beta''J''}
    \dipsq{\beta''J''}{\beta J}}
    {\left[(E_{\beta''J''}-E_{\beta J})^2 
    -\hbar^2\omega^2\right]^2}
  \label{eq:alpha-vect-im} \\
  \Re[\alpha_\mathrm{tens}(\omega)] & = -\sum_{\beta''J''}
   \frac{3[J''(J''+1)-J(J+1)]^2-9J''(J''+1)+J(J+1)+6}
    {3(J+1)(2J+1)(2J+3)} \times
   \frac{ (E_{\beta''J''}-E_{\beta J})
     \dipsq{\beta''J''}{\beta J}}
    {(E_{\beta''J''}-E_{\beta J})^2-\hbar^2\omega^2}
  \label{eq:alpha-tens-re} \\
  \Im[\alpha_\mathrm{tens}(\omega)] & = -\sum_{\beta''J''} 
   \frac{3[J''(J''+1)-J(J+1)]^2-9J''(J''+1)+J(J+1)+6}
    {6(J+1)(2J+1)(2J+3)}
  \nonumber \\
   & \times \frac{ \hbar\gamma_{\beta''J''}
     \left[(E_{\beta''J''}-E_{\beta J})^2+\hbar^2\omega^2\right]
     \dipsq{\beta''J''}{\beta J}}
    {\left[(E_{\beta''J''}-E_{\beta J})^2 
     -\hbar^2\omega^2\right]^2}
  \label{eq:alpha-tens-im}
\end{align}
\end{widetext}
Note that in Eqs.~(7), (8) and (11) of Ref.~\cite{lepers2014}, the sign of the vector polarizabiity is not correct; the error has been fixed in Eqs.~(\ref{eq:alpha-vect-re}) and (\ref{eq:alpha-vect-im}) above.

\subsection{Effect of a single intermediate configuration}
\label{sub:1array}

In this subsection, we assume that the intermediate levels $|\beta'' J'' \rangle$ appearing in Eq.~\eqref{eq:alpha-k} all belong to the same configuration, and that their transition energies $E_{\beta''J''}-E_{\beta J}$ can be replaced by a single effective one. Moreover, we assume that the configurations of the $|\beta J\rangle$ and $|\beta'' J'' \rangle$ levels differ by the hopping of only one valence electron; in other words, we ignore transitions involving the $4f$ core electrons. This will yield analytical expressions useful to estimate $\alpha_k(\omega)$, and to understand the trapping in some relevant levels, like those belonging to the lowest or the [Xe]$\,4f^q6s6p$ configurations.

Many levels of lanthanide atoms can be interpreted in the frame of the $jj$ coupling scheme. The electronic core, containing the $4f$ shell, is characterized by its orbital $L_c$, spin $S_c$ and total electronic angular momentum $J_c$. The valence electrons belong for instance to the $5d$, $6s$ or $6p$ shells. This group of electrons is characterized by their orbital $L_v$, spin $S_v$ and total electronic angular momentum $J_v$. Then $J_c$ and $J_v$ are coupled to give the total electronic angular momentum $J$ of the atomic level. In the present study, we focus on configurations $[\mathrm{Xe}]\, 4f^q.n_1\ell_1n_2\ell_2$ ($q=10$, 11, 12 for Dy, Ho, Er, respectively) with two valence electrons, including \textit{e.g.}~$4f^q6s^2$ or $4f^q6s6p$; but our results can be extended to configurations with 3 valence electrons like $4f^{q-1}5d6s^2$ or $4f^{q-1}5d^26s$. The full label of the level is therefore $[\mathrm{Xe}]\,4f^q(^{2S_c+1}L_c{}_{J_c}) . n_1\ell_1n_2\ell_2 (^{2S_v+1}L_v{}_{J_v}) \, (J_c,J_v)_J$, and its electronic parity is $(-1)^{q+\ell_1+\ell_2}$. In what follows, we will omit the xenon core [Xe] in electronic configurations.

It is worthwhile to note that the levels of the $4f^q5d6s$ configuration are better described in the $jK$ coupling scheme $^{2S_v+1}[K]_J$: $J_c$ is firstly coupled with $L_v$ to give $K$, itself coupled with $S_v$ to give $J$. In order to calculate the polarizability of such levels, it is necessary to apply the basis transformation from $jj$ to $jK$ coupling schemes \cite{cowan1981}. However, if those levels appear in the sum over $|\beta''J''\rangle$, the $jj$ coupling scheme is sufficient, as all the levels of the $4f^q5d6s$ configuration are assumed to have the same energy (see paragraph \ref{ssub:1ar-real}).

\subsubsection{Transition dipole moment in $jj$ coupling}

In the electric-dipole (E1) approximation, the transitions with the strongest dipole moments are those for which one valence electron, say $n_2\ell_2$, is promoted to an orbital $n''_2\ell''_2$ such that $\ell''_2=\ell_2\pm 1$. The angular momenta of the atom must also satisfy the selection rules:  $L''_v=L_v$ or $L_v\pm 1$, $S''_v=S_v$, $J''_v =J_v$ or $J_v\pm 1$, and $J''=J$ or $J\pm 1$, excluding transitions between couples of angular momenta $(0,0)$, whereas the quantum numbers of the core are not modified ($L''_c=L_c$, $S''_c=S_c$ and $J''_c=J_c$). In the frame of the $jj$ coupling scheme, we can express the reduced transition dipole moment between the levels $|\beta J\rangle$ and $|\beta''J''\rangle$ as a function of the mono-electronic transition dipole moment $\langle n''_2\ell''_2 |\hat{r}|n_2\ell_2 \rangle$ expressed as the matrix element of the mono-electronic $\hat{r}$-operator. We apply the following successive steps \cite{cowan1981}.

By writing atomic levels as the lists of quantum numbers $|n_1\ell_1 n_2\ell_2 L_v S_v J_v J_c J\rangle$ (and similarly for double-primed quantum numbers), we start working with ($J_c$, $J_v$, $J$),
\begin{align}
  & \dipsq{n_1\ell_1 n''_2\ell''_2 L''_v S_v J''_v J_c J''}
    {n_1\ell_1 n_2\ell_2 L_v S_v J_v J_c J} 
  \nonumber \\
  & = \left(2J+1\right) \left(2J''+1\right)
    \sixj{J_v}{J_c}{J}{J''}{1}{J''_v}^2
  \nonumber \\
  & \times \dipsq{n_1\ell_1 n''_2\ell''_2 L''_v S_v J''_v}
    {n_1\ell_1 n_2\ell_2 L_v S_v J_v} \,.
  \label{eq:dipsq-1}
\end{align}
Then we go one step further with ($L_v$, $S_v$, $J_v$)
\begin{align}
  & \dipsq{n_1\ell_1 n''_2\ell''_2 L''_v S_v J''_v}
    {n_1\ell_1 n_2\ell_2 L_v S_v J_v} 
  \nonumber \\
  & = \left(2J_v+1\right) \left(2J''_v+1\right)
    \sixj{L_v}{S_v}{J_v}{J''_v}{1}{L''_v}^2 
  \nonumber \\
    & \times \dipsq{n_1\ell_1 n''_2\ell''_2 L''_v}
    {n_1\ell_1 n_2\ell_2 L_v} \,,
  \label{eq:dipsq-2}
\end{align}
and with ($n_1$, $\ell_1$, $n_2$, $\ell_2$, $L_v$),
\begin{align}
  & \dipsq{n_1\ell_1 n''_2\ell''_2 L''_v}
            {n_1\ell_1 n_2  \ell_2   L_v}  
  \nonumber \\
  & = \left(1 + \delta_{n_1n_2} \delta_{\ell_1\ell_2} \right)
    \left(1 + \delta_{n_1n''_2} \delta_{\ell_1\ell''_2} \right)
    \left(2L_v+1\right)
  \nonumber \\
  & \times \left(2L''_v+1\right)
    \sixj{\ell_2}{\ell_1}{L_v}{L''_v}{1}{\ell''_2}^2
    \dipsq{n''_2\ell''_2}{n_2\ell_2} \,,
  \label{eq:dipsq-3}
\end{align}
where the $\delta$'s are Kronecker symbols, which bring a factor of 2 for equivalent electrons $(n_1\ell_1)=(n_2\ell_2)$ or $(n_1\ell_1)=(n''_2\ell''_2)$. Finally,
\begin{align}
  \dipsq{n''_2\ell''_2}{n_2\ell_2} & = 
    e^2 r_{n_2\ell_2,n''_2\ell''_2}^2
    \left(2\ell_2+1\right) \nonumber \\
    & \times \left(2\ell''_2+1\right)
    \thrj{\ell''_2}{1}{\ell_2}{0}{0}{0}^2\,,
  \label{eq:dipsq-4}
\end{align}
where $(:::)$ is a Wigner 3-j symbol, $e$ the absolute value of the electronic charge, and $r_{n_2\ell_2,n''_2\ell''_2} = r_{n''_2\ell''_2,n_2\ell_2} \equiv \langle n''_2\ell''_2|\hat{r}|n_2\ell_2\rangle$.

\subsubsection{Real part of the polarizability}
\label{ssub:1ar-real}

We assume that the polarizability $\alpha_k(\omega)$ of the level $|\beta J\rangle$, see Eq.~(\ref{eq:alpha-k}), involves transitions towards levels $|\beta'' J''\rangle$ belonging to configurations of the kind $4f^q.n_1\ell_1.n''_2\ell''_2$. By separating the contributions of those configurations, we can write
\begin{equation}
  \alpha_k(\omega) = \sum_{n''_2\ell''_2}
    \alpha_k^{n''_2\ell''_2}(\omega),
  \label{eq:alpha-k-sum}
\end{equation}
which relies on two main hypothesis: (i) Transitions to levels of configurations in which one core electron is excited, \textit{e.g.}~$4f^{q-1}.5d.n_1\ell_1.n_2\ell_2$	 are excluded, as they are often significantly weaker. (ii) Configuration interaction (CI) is totally neglected, both between different configurations of the kind $4f^q.n_1\ell_1.n''_2\ell''_2$, and with those of the kind $4f^{q-1}.n''\ell''.n_1\ell_1.n_2\ell_2$. The effect of CI will be addressed in the next subsection.

The central assumption of this work is that the energy differences implying the levels of a given configuration can be replaced by a single effective energy $\hbar \omega_{n''_2\ell''_2}$,
\begin{equation}
  E_{\beta''J''} - E_{\beta J} \approx \hbar
    \omega_{n''_2\ell''_2} \,.
  \label{eq:omega-eff}
\end{equation}
The validity of this assumption depends on the frequency $\omega$ at which the DDPs are calculated, which should not ``fall'' into the levels of the $4f^q.n_1\ell_1.n''_2\ell''_2$ configuration. If we denote $\min(E_{\beta''J''})$ and $\max(E_{\beta''J''})$ their smallest and largest energies, equation \eqref{eq:omega-eff} is not applicable for
\begin{equation}
  \min(E_{\beta''J''}) - E_{\beta J} \lesssim \bar{\omega} 
    \lesssim \max(E_{\beta''J''}) - E_{\beta J}
\end{equation}
where $\bar{\omega} = \pm \omega$ for $E_{\beta''J''} > E_{\beta J}$ and $E_{\beta''J''} < E_{\beta J}$ respectively.
For ground-level Ho, the excluded frequencies, which correspond to the energies of the $4f^{11}(^4I^o_{15/2}) . 6s6p(^1P^o_1) \, (15/2,1)$ manifold, roughly range from 23000 to 24000 cm$^{-1}$.

By consequence the sum in Eq.~(\ref{eq:alpha-k}) is restricted to the quantum numbers $L''_v$, $J''_v$ and $J''$ allowed by electric-dipole transitions. (For configurations with at least one $s$ electron, there is obviously only one possible $L_v$ value.) Inserting Eq.~(\ref{eq:alpha-k-sum}) into Eq.~(\ref{eq:alpha-k-re}), we can extract the real part
\begin{align}
  & \Re[\alpha_k^{n''_2\ell''_2}(\omega)]
  \nonumber \\
  & = \frac{2\left( \omega_{n''_2\ell''_2} \, \delta_{(-1)^k,1}
       - \omega \, \delta_{(-1)^k,-1} \right)}
    {\hbar\left(\omega_{n''_2\ell''_2}^2-\omega^2\right)}
  \nonumber \\
  & \times \sqrt{2k+1} \sum_{L''_vJ''_vJ''} (-1)^{J+J''}
   \sixj{1}{1}{k}{J}{J}{J''}
  \nonumber \\
  & \times \dipsq{n_1\ell_1 n''_2\ell''_2 L''_v S_v J''_v J_c J''}
    {n_1\ell_1 n_2\ell_2 L_v S_v J_v J_c J} \,.
  \label{eq:alpha-k-pp-1}
\end{align}
Using Eq.~(\ref{eq:dipsq-1}), we obtain
\begin{align}
  & \Re[\alpha_k^{n''_2\ell''_2}(\omega)] 
  \nonumber \\
  & = \frac{2\left( \omega_{n''_2\ell''_2} 
    \times \delta_{(-1)^k,1}
     - \omega \times \delta_{(-1)^k,-1} \right)}
    {\hbar\left(\omega_{n''_2\ell''_2}^2-\omega^2\right)}
  \nonumber \\
  & \times \sqrt{2k+1} \sum_{L''_vJ''_vJ''} (-1)^{J+J''}
   \sixj{1}{1}{k}{J}{J}{J''}
  \nonumber \\
  & \times \left(2J+1\right) \left(2J''+1\right)
   \sixj{J_v}{J_c}{J}{J''}{1}{J''_v}^2
  \nonumber \\
  & \times \dipsq{n_1\ell_1 n''_2\ell''_2 L''_v S_v J''_v}
    {n_1\ell_1 n_2\ell_2 L_v S_v J_v} \,.
  \label{eq:alpha-k-pp-2}
\end{align}
To calculate this expression, we note that the quantum number $J''$ only appears in angular terms, so that we use the identity (see Ref.~\cite{varshalovich1988}, p.~305)
\begin{align}
  & \sum_X \left(-1\right)^{R+X} \left(2X+1\right)
   \sixj{a}{b}{p}{c}{d}{X} \sixj{c}{d}{q}{e}{f}{X}
  \nonumber \\
  & \; \times \sixj{e}{f}{r}{b}{a}{X}
   = \sixj{p}{q}{r}{e}{a}{d} \sixj{p}{q}{r}{f}{b}{c}
  \label{eq:3-6j}
\end{align}
with $R = a+b+c + d+e+f + p+q+r$, as well as the invariance properties of Wigner 6-j symbols with respect to line and column permutations. Applying Eq.~\eqref{eq:3-6j} with $a=d=1$, $b=c=J$, $e=J''_v$, $f=J_c$, $p=k$ and $q=r=J_v$, we can get rid of $J''$ in Eq.~\eqref{eq:alpha-k-pp-2}
\begin{align}
  & \Re[\alpha_k^{n''_2\ell''_2}(\omega)] 
  \nonumber \\
  & = \frac{2\left( \omega_{n''_2\ell''_2} \, \delta_{(-1)^k,1}
     - \omega \, \delta_{(-1)^k,-1} \right)}
    {\hbar\left(\omega_{n''_2\ell''_2}^2-\omega^2\right)}
  \nonumber \\
  & \times \sqrt{2k+1} \sum_{L''_vJ''_v}
   (-1)^{J_c+2J_v+J''_v+J+k}
  \nonumber \\
  & \times \left(2J+1\right) 
   \sixj{1}{1}{k}{J_v}{J_v}{J''_v}
   \sixj{J_v}{J_c}{J}{J}{k}{J_v}
  \nonumber \\
  & \times \dipsq{n_1\ell_1 n''_2\ell''_2 L''_v S_v J''_v}
    {n_1\ell_1 n_2\ell_2 L_v S_v J_v} \,.
  \label{eq:alpha-k-pp-3}
\end{align}

At this point, it is worthwhile to note the following fact \cite{angel1968}. The definitions of the coupled polarizabilities $\alpha_k(\omega)$ and $\alpha_k^{n''_2\ell''_2}(\omega)$, given respectively by Eqs.~\eqref{eq:alpha-k} and \eqref{eq:alpha-k-sum}, are such that they can be written as the reduced matrix elements of the operators $\hat{\alpha}_k(\omega)$ and $\hat{\alpha}_k ^{n''_2\ell''_2}(\omega)$, which are tensors of rank $k$. In particular, one can resort to the Wigner-Eckart theorem \cite{varshalovich1988} to calculate the coupled polarizability of a level $|\beta JM\rangle$, namely $\langle\beta JM| \hat{\alpha}_k(\omega) |\beta JM\rangle = \langle\beta J\Vert \hat{\alpha}_k(\omega) \Vert\beta J\rangle \times C_{JMk0}^{JM} / \sqrt{2J+1}$, with $C_{JMk0}^{JM}$ a Clebsh-Gordan coefficient (and similarly for $\hat{\alpha}_k ^{n''_2\ell''_2}(\omega)$). One can also apply the transformation of tensor operators regarding angular-momentum basis sets; in this respect, equation \eqref{eq:alpha-k-pp-3} can be seen as such a transformation,
\begin{align}
  & \left\langle n_1\ell_1 n_2\ell_2 L_v S_v J_v J_c J 
   \right\Vert \Re[\hat{\alpha}_k^{n''_2\ell''_2}(\omega)]
   \left\Vert n_1\ell_1 n_2\ell_2 L_v S_v J_v J_c J \right\rangle
  \nonumber \\
  & = \left(-1\right)^{J_c+J_v+k+J} \left(2J+1\right)
   \sixj{J_v}{J_c}{J}{J}{k}{J_v}
  \nonumber \\
  & \times \left\langle n_1\ell_1 n_2\ell_2 L_v S_v J_v
    \right\Vert \Re[\hat{\alpha}_k^{n''_2\ell''_2}(\omega)]
   \left\Vert n_1\ell_1 n_2\ell_2 L_v S_v J_v \right\rangle 
  \label{eq:alpha-k-transfo} \,,
\end{align}
where the last line reads
\begin{align}
  & \left\langle n_1\ell_1 n_2\ell_2 L_v S_v J_v
    \right\Vert \Re[\hat{\alpha}_k^{n''_2\ell''_2}(\omega)]
   \left\Vert n_1\ell_1 n_2\ell_2 L_v S_v J_v \right\rangle
  \nonumber \\
  & = \frac{2\left( \omega_{n''_2\ell''_2} 
    \times \delta_{(-1)^k,1}
     - \omega \times \delta_{(-1)^k,-1} \right)}
    {\hbar\left(\omega_{n''_2\ell''_2}^2-\omega^2\right)}
  \nonumber \\
  & \times \sqrt{2k+1} \sum_{J''_vL''_v} (-1)^{J_v+J''_v}
    \sixj{1}{1}{k}{J_v}{J_v}{J''_v}
  \nonumber \\
   & \times \dipsq{n_1\ell_1 n''_2\ell''_2 L''_v S_v J''_v}
    {n_1\ell_1 n_2\ell_2 L_v S_v J_v} \,.
  \label{eq:alpha-k-transfo-2}
\end{align}

Coming back to our main purpose, we apply equation (\ref{eq:3-6j}) twice more: firstly with Eq.~(\ref{eq:dipsq-2}) to express the sum over $J''_v$, and secondly with Eq.~(\ref{eq:dipsq-3}) to express the sum over $L''_v$. Doing so, we get to the final expression
\begin{widetext}
\begin{eqnarray}
  \Re[\alpha_k^{n''_2\ell''_2}(\omega)] & = &
   \frac{2\sqrt{2k+1}}{\hbar} \times
   \frac{\omega_{n''_2\ell''_2} \, \delta_{(-1)^k,1}
     - \omega \, \delta_{(-1)^k,-1}}
    {\omega_{n''_2\ell''_2}^2-\omega^2}
   \left(1 + \delta_{n_1n_2} \delta_{\ell_1\ell_2} \right)
   \left(1 + \delta_{n_1n''_2} \delta_{\ell_1\ell''_2} \right)
  \nonumber \\
  & \times & (-1)^{J+J_c-S_v+\ell_1+\ell''_2+k}
   \left(2J+1\right) \left(2J_v+1\right) \left(2L_v+1\right)
   \left(2\ell_2+1\right) \left(2\ell''_2+1\right)
   \sixj{J_v}{J_c}{J}{J}{k}{J_v}
  \nonumber \\
  & \times & \sixj{L_v}{S_v}{J_v}{J_v}{k}{L_v}
   \sixj{\ell_2}{\ell_1}{L_v}{L_v}{k}{\ell_2}
   \sixj{1}{1}{k}{\ell_2}{\ell_2}{\ell''_2}
   \thrj{\ell''_2}{1}{\ell_2}{0}{0}{0}^2
   e^2 r_{n_2\ell_2,n''_2\ell''_2}^2
  \label{eq:alpha-k-pp-4}
\end{eqnarray}
\end{widetext}
which depends on two effective parameters: the transition frequency $\omega_{n''_2\ell''_2}$ and the mono-electronic transition dipole moment $-er_{n_2\ell_2,n''_2\ell''_2}$.

The rest of Eq.~\eqref{eq:alpha-k-pp-4} consists in very insightful angular terms. In particular, the 6-j symbols indicate that, if one of the quantum numbers $J$, $J_v$ or $L_v$ is equal to 0, then the vector and tensor polarizabilities, proportional to $\Re[\alpha_{k=1}^{n''_2\ell''_2}(\omega)]$ and $\Re[\alpha_{k=2}^{n''_2\ell''_2}(\omega)]$) respectively, vanish. This is for instance the case for lanthanides in their ground level, which is characterized by $L_v=J_v=0$. In our full numerical calculation of the polarizability \cite{lepers2014, li2017}, we have shown that indeed the vector and tensor contributions are much weaker than the scalar one. Equation (\ref{eq:alpha-k-pp-4}) tends to confirm that those weak contributions come from transitions in which one $4f$ electron is excited. Such conclusions are also valid for any level belonging to the lowest configuration $4f^q6s^2$, as shown in our previous articles (see Ref.~\cite{li2017} and in subsection \ref{sub:pola}).

\subsubsection{Imaginary part of the polarizability}

For the imaginary part to be relevant, we consider a metastable level $|\beta J\rangle$,\idest{whose}natural linewidth $\gamma_{\beta J}$ is negligible compared to the photon-scattering rate induced by the electromagnetic field \cite{grimm2000, lepers2014}. In practice, this may concern excited levels of the lowest configuration $4f^q6s^2$ or the levels $4f^q(^{2S_c+1}L_c{}_{J_c}) . 6s6p(^3P_2) \, (J_c,2)_{J_c+2}$, which have no decay channel in the E1 approximation (except for the level $(6,2)_8^o$ of Er) \cite{NIST_ASD}.

As Eq.~\eqref{eq:alpha-k-im} shows, the imaginary part of the polarizability involves the natural linewidth of intermediate levels $|\beta''J''\rangle$,
\begin{align}
  \gamma_{\beta''J''} & = \sum_{ \tilde{\beta}\tilde{J} ,\;
   E_{\tilde{\beta}\tilde{J}} < E_{\beta''J''}}
   A_{\beta''J'',\tilde{\beta}\tilde{J}}
  \nonumber \\
   & = \frac{\sum_{\tilde{\beta}\tilde{J}}
    (E_{\beta''J''}-E_{\tilde{\beta}\tilde{J}})^3
    |\langle \beta''J'' \Vert \mathbf{d}
     \Vert \tilde{\beta}\tilde{J} \rangle |^2}
   {3\pi \epsilon_0 \hbar^4 c^3 (2J''+1)}
  \label{eq:gamma-nat}
\end{align}
where $A_{\beta''J'',\tilde{\beta}\tilde{J}}$ is the transition probability characterizing the spontaneous emission from the level $|\beta''J''\rangle$ to the level $|\tilde{\beta}\tilde{J}\rangle$. We focus on the influence of the $\beta''J''$ levels belonging to the configuration $4f^q.n_1\ell_1.n''_2\ell''_2$. In addition, we assume that the latter levels only decay towards levels $|\tilde{\beta} \tilde{J} \rangle$ belonging to the configuration $4f^q. n_1\ell_1. n_2\ell_2$. Therefore the sum in Eq.~\eqref{eq:gamma-nat} runs over the quantum numbers $\tilde{J}$, $\tilde{J}_v$ and $\tilde{L}_v$. If we express the squared reduced transition dipole moment as in Eq.~\eqref{eq:dipsq-1}, Equation \eqref{eq:gamma-nat} becomes
\begin{align}
  \gamma_{L''_vJ''_vJ''} & = \frac{\omega_{n''_2\ell''_2}^3}
   {3\pi \epsilon_0 \hbar c^3}
   \sum_{\tilde{L}_v \tilde{J}_v \tilde{J}} (2\tilde{J}+1)
   \sixj{\tilde{J}_v}{J_c}{\tilde{J}}{J''}{1}{J''_v}^2
  \nonumber \\
   & \times \dipsq{n_1\ell_1 n''_2\ell''_2 L''_vS_vJ''_v}
   {n_1\ell_1 n_2\ell_2 \tilde{L_v} S_v \tilde{J}_v}
  \label{eq:gamma-nat-2}
\end{align}
Since $\tilde{J}$ only appears in angular factors, the sum over $\tilde{J}$ reduces to the orthogonalization relations of 6-j symbols,
\begin{equation}
  \sum_{\tilde{J}} (2\tilde{J}+1)
   \sixj{\tilde{J}_v}{J_c}{\tilde{J}}{J''}{1}{J''_v}^2
   = \frac{1}{2J''_v+1} \,.
  \label{eq:orth-6j}
\end{equation}
By using Eqs.~\eqref{eq:dipsq-2} and \eqref{eq:dipsq-3} for the transition dipole moment, we can calculate the sums over $\tilde{J}_v$ and $\tilde{L}_v$ in a similar way, and finally we get to the expression [see also Eq.~\eqref{eq:dipsq-4}]
\begin{eqnarray}
  \gamma_{L''_vJ''_vJ''} & = & \frac{\omega_{n''_2\ell''_2}^3
   r_{n_2\ell_2,n''_2\ell''_2}^2}
   {3\pi \epsilon_0 \hbar c^3}
   \left(2\ell_2+1\right) \thrj{\ell''_2}{1}{\ell_2}{0}{0}{0}^2
  \nonumber \\
    & \times & 
   \left(1 + \delta_{n_1n_2} \delta_{\ell_1\ell_2} \right)
   \left(1 + \delta_{n_1n''_2} \delta_{\ell_1\ell''_2} \right)\,.
  \label{eq:gamma-nat-3}
\end{eqnarray}

Strikingly the natural linewidth of the intermediate levels does not depend on $L''_v$, $J''_v$ or $J''$; it is identical for all the levels of the $4f^q.n_1\ell_1.n''_2\ell''_2$ configuration. In calculating $\Im[\alpha_k^{n''_2\ell''_2}(\omega)]$, we can factorize $\gamma_{L''_vJ''_vJ''}$ out of the sum over $L''_v$, $J''_v$ and $J''$, and so the similar steps [see Eqs.~\eqref{eq:alpha-k-pp-2}--\eqref{eq:alpha-k-pp-4}] as for the real part can be applied, which leads to the final expression
\begin{widetext}
\begin{eqnarray}
  \Im[\alpha_k^{n''_2\ell''_2}(\omega)] & = &
   \frac{\omega_{n''_2\ell''_2}^3\sqrt{2k+1}}
    {3\pi \epsilon_0 \hbar^2 c^3} \times
   \frac{(\omega_{n''_2\ell''_2}^2+\omega^2)\, \delta_{(-1)^k,1}
     - 2\omega\omega_{n''_2\ell''_2} \, \delta_{(-1)^k,-1}}
    {(\omega_{n''_2\ell''_2}^2-\omega^2)^2}
   \left(1 + \delta_{n_1n_2} \delta_{\ell_1\ell_2} \right)^2
   \left(1 + \delta_{n_1n''_2} \delta_{\ell_1\ell''_2} \right)^2
  \nonumber \\
  & \times & (-1)^{J+J_c-S_v+\ell_1+\ell''_2+k}
   \left(2J+1\right) \left(2J_v+1\right) \left(2L_v+1\right)
   \left(2\ell_2+1\right)^2 \left(2\ell''_2+1\right)
   \sixj{J_v}{J_c}{J}{J}{k}{J_v}
  \nonumber \\
  & \times & \sixj{L_v}{S_v}{J_v}{J_v}{k}{L_v}
   \sixj{\ell_2}{\ell_1}{L_v}{L_v}{k}{\ell_2}
   \sixj{1}{1}{k}{\ell_2}{\ell_2}{\ell''_2}
   \thrj{\ell''_2}{1}{\ell_2}{0}{0}{0}^4
   e^4 r_{n_2\ell_2,n''_2\ell''_2}^4 \,.
  \label{eq:alpha-k-pp-5}
\end{eqnarray}
\end{widetext}
Therefore, similarly to the real part, the imaginary part of the polarizability depends on the effective frequency $\omega_{n''_2\ell''_2}$ and the mono-electronic transition dipole moment $-er_{n_2\ell_2,n''_2\ell''_2}$ between the two configuration, and on some angular factors.

Again, those angular factors show that, if one of the quantum numbers $J$, $J_v$ or $L_v$ is equal to zero, then the vector and tensor polarizabilities are equal to zero. For lanthanide atoms in the ground level, our simplified model predicts that both the real and imaginary parts of the vector and tensor DDPs vanish, see resp.~Eqs.~\eqref{eq:alpha-k-pp-4} and \eqref{eq:alpha-k-pp-5}. For the real part, that prediction agrees with our full numerical calculation \cite{lepers2014, li2017} (see also subsection \ref{sub:pola}); but the imaginary part it does not. To explain that contradiction, we note that the vector-to-scalar and tensor-to-scalar ratios are significantly larger for Er than for Dy and Ho. In addition, Er is the only atom among the three for which we modeled the excited levels including the configurations $4f^{11}5d6s^2$, $4f^{11}5d^26s$ and $4f^{12}6s6p$, and so for which we expect to have a better description of CI mixing. This tends to prove that CI plays an important role in the imaginary part of the DDPs. That is why, in the next subsection, we will improve our model by taking into account CI among excited levels.

\subsection{Effect of configuration interaction}
\label{sub:ci}

We focus on the polarizability of the ground level of lanthanides, denoted $|\beta J\rangle \equiv |0J\rangle$, and characterized by $L_v=S_v=0$ and $J=J_c$. According to Eqs.~\eqref{eq:alpha-k-pp-4} and \eqref{eq:alpha-k-pp-5}, there are three excited levels, denoted $|1J''\rangle$ for $J''=J$ and $J\pm 1$, which contribute to the polarizability; they are characterized by $L''_v=J''_v=1$. In this section, we consider that these levels can be mixed by CI to other levels $|mJ''\rangle$ belonging to other configurations. Therefore the eigenvector of the excited levels $|\beta''J''\rangle$ can be expanded as
\begin{equation}
  \left|\beta''J''\right\rangle = \sum_{m\ge 1}
   c_{\beta''m}^{(J'')} \left|mJ''\right\rangle \,,
  \label{eq:ci-wf}
\end{equation}
where $|mJ''\rangle$ are from now called basis states. Furthermore we assume that the state $|1J''\rangle$ is the only one contributing to the transition dipole moment $\dip{0J}{\beta''J''}$. This is exactly valid for basis states of the $4f^{q-1}5d^26s$ configuration, and approximately valid for states of the $4f^{q-1}5d6s^2$ configuration, as the latter contribute significantly less than the states of the $4f^q6s6p$ configuration.

In this case the squared transition dipole moment reads
\begin{align}
  \dipsq{\beta''J''}{0J} & = \left|c_{\beta'',1}^{(J'')}\right|^2
   \dipsq{1J''}{0J}
  \notag \\
   & = \frac{2}{3} w_{\beta'',1}^{(J'')} \,
   (2J''+1) r_{6s6p}^2
  \label{eq:dipsq-gs}
\end{align}
where, in the second line, we expressed $\dipsq{1J''}{0J}$ using Eqs.~\eqref{eq:dipsq-1}--\eqref{eq:dipsq-4} and the explicit forms of the 3-j and 6-j symbols. In Eq.~\eqref{eq:dipsq-gs} we introduced the weights $w_{\beta'',1}^{(J'')} = \left|c_{\beta'',1}^{(J'')}\right|^2$ of the basis states $|1J''\rangle$ in the levels $|\beta''J''\rangle$, which satisfy the normalization conditions
\begin{equation}
  \sum_{m\ge 1} w_{\beta''m}^{(J'')} 
   = \sum_{\beta''} w_{\beta''m}^{(J'')} = 1
  \label{eq:orth-w}
\end{equation}
for each $J''$ separately.

Turning to the polarizability, we find that the real part is
\begin{align}
  & \Re[\alpha_k^{6p}(\omega)] \nonumber \\
  & = \frac{4e^2 r_{6s6p}^2 \sqrt{2k+1}}{3\hbar} \,
   \frac{\omega_{6p} \, \delta_{(-1)^k,1}
     - \omega \, \delta_{(-1)^k,-1}}
    {\omega_{6p}^2-\omega^2}
  \nonumber \\
  & \times \sum_{J''} \left(-1\right)^{J+J''} 
   (2J''+1) \sixj{1}{1}{k}{J}{J}{J''}
   \sum_{\beta''} w_{\beta'',1}^{(J'')}
  \nonumber \\
  & = -\frac{4e^2 r_{6s6p}^2 \omega_{6p} \, \delta_{k0}}
   {\hbar (\omega_{6p}^2 - \omega^2)}
   \sqrt{\frac{2J+1}{3}}
  \label{eq:alpha-ci-1}
\end{align}
where the sums over $\beta''$ and $J''$ are calculated using respectively Eq.~\eqref{eq:orth-w} and (see Ref.~\cite{varshalovich1988}, p.~305)
\begin{align}
  & \sum_X (-1)^{a+b+X} (2X+1) \sixj{a}{a}{c}{b}{b}{X}
  \nonumber \\
  & \quad = \delta_{c0} \sqrt{(2a+1)(2b+1)}
  \label{eq:sum-6j}
\end{align}
with $a=1$, $b=J$, $c=k$ and $X=J''$. Equation \eqref{eq:alpha-ci-1} shows that the vector and tensor polarizabilities vanish for lanthanides in the ground level (or in any level of the electronic configuration $4f^q6s^2$), whatever the CI mixing in the excited levels. In this respect the inclusion of CI in our model does not modify the conclusions of the single-configuration case [see Eq.~\eqref{eq:alpha-k-pp-4}]: the only contribution is the scalar one $\Re[\alpha_\mathrm{scal}(\omega)] = 4e^2 r_{6s6p}^2 \omega_{6p} /
3\hbar (\omega_{6p}^2 - \omega^2)$.

In order to calculate the imaginary part of the polarizability, we recall that the excited level $|1J''\rangle$ can only decay toward the ground level $|0J\rangle$. Therefore $\Im[\alpha_k^{6p}(\omega)]$ reads
\begin{align}
  \Im[\alpha_k^{6p}(\omega)] & =
   \frac{(\omega_{6p}^2+\omega^2)\, \delta_{(-1)^k,1}
     - 2\omega\omega_{6p} \, \delta_{(-1)^k,-1}}
    {(\omega_{6p}^2-\omega^2)^2}
  \nonumber \\
   & \times \frac{\omega_{6p}^3\sqrt{2k+1}}
                 {3\pi \epsilon_0 \hbar^2 c^3} 
   \sum_{J''} \left(-1\right)^{J+J''} \sixj{1}{1}{k}{J}{J}{J''}
  \nonumber \\
   & \times \sum_{\beta''} \frac{1}{2J''+1}
   \left| \dip{\beta''J''}{0J} \right|^4
  \nonumber \\
  & = \frac{(\omega_{6p}^2+\omega^2)\, \delta_{(-1)^k,1}
     - 2\omega\omega_{6p} \, \delta_{(-1)^k,-1}}
    {(\omega_{6p}^2-\omega^2)^2}
  \nonumber \\
   & \times \frac{4\omega_{6p}^3 r_{6s6p}^4 \sqrt{2k+1}}
                 {27\pi \epsilon_0 \hbar^2 c^3} 
   \sum_{J''} \left(-1\right)^{J+J''} \left(2J''+1\right)
  \nonumber \\
   & \times \sixj{1}{1}{k}{J}{J}{J''}
   \sum_{\beta''} \left( w_{\beta'',1}^{(J'')} \right)^2
  \label{eq:alpha-ci-2}
\end{align}
where we took the square of Eq.~\eqref{eq:dipsq-gs}.

Equation \eqref{eq:alpha-ci-2} is a key result of this work. 
Contrary to the real part given by Eq.~\eqref{eq:alpha-ci-1}, the sum over $\beta''$ cannot be simplified in the imaginary part of the polarizability, as it involves the squared weights of the $|1J''\rangle$ basis vectors in the excited levels $|\beta''J''\rangle$. In this respect, we can say that the imaginary part of the polarizability is more sensitive to the details of the atomic structure than the real part.

In particular, taking the square of Eq.~\eqref{eq:orth-w}, we find that
\begin{align}
  \sum_{\beta''} \left( w_{\beta''m}^{(J'')} \right)^2
   & = \left( \sum_{\beta''} w_{\beta''m}^{(J'')} \right)^2
   - 2 \sum_{\substack{\beta''_1\beta''_2 \\
    \beta''_1 < \beta''_2}}
   w_{\beta''_1m}^{(J'')} w_{\beta''_2m}^{(J'')}
  \nonumber \\
   & = 1 - 2 \sum_{\substack{\beta''_1\beta''_2 \\
    \beta''_1 < \beta''_2}}
   w_{\beta''_1m}^{(J'')} w_{\beta''_2m}^{(J'')} \le 1
  \label{eq:sum-w-sq}
\end{align}
where $\beta''_1 < \beta''_2$ means $E_{\beta''_1J''} < E_{\beta''_2J''}$, to avoid double counting. The inequality comes from the fact that $w_{\beta''m}^{(J'')} \ge 0$, $\forall\,m$, $\beta''$, $J''$. The limit for which equation \eqref{eq:sum-w-sq} is unity corresponds to the case where one weight is unity and all the others are zero, \textit{i.e.}~no CI. In this particular case, the sums over $\beta''$ and $J''$ in Eq.~\eqref{eq:alpha-ci-2} can be simplified,
\begin{equation}
  \Im[\alpha_k^{6p}(\omega)] = -\frac{4\omega_{6p}^3
   (\omega_{6p}^2+\omega^2) e^4 r_{6s6p}^4 \delta_{k0}}
   {9\pi \epsilon_0 \hbar^2 c^3 (\omega_{6p}^2-\omega^2)^2} 
   \sqrt{\frac{2J+1}{3}}
  \label{eq:alpha-ci-3}
\end{equation}
and so $\Im[\alpha_\mathrm{scal}^{6p}(\omega)] = 4\omega_{6p}^3 (\omega_{6p}^2 + \omega^2) e^4 r_{6s6p}^4 / 27\pi \epsilon_0 \hbar^2 c^3$ $(\omega_{6p}^2-\omega^2)^2$, which can also be obtained from Eq.~\eqref{eq:alpha-k-pp-5}.
By comparing Eqs.~\eqref{eq:alpha-ci-2}--\eqref{eq:alpha-ci-3}, we find that CI has two effects:
\begin{itemize}
  \item It tends to reduce the scalar contribution $\Im[\alpha_0^{6p}(\omega)]$. Indeed in the limit of strong CI mixing, when $N$ basis states $|mJ''\rangle$ ($m=1$ to $N$) are equally spread over $N$ excited levels $|\beta''J''\rangle$, which means $w_{\beta''m}^{(J'')} = 1/N$ for all $J''$, then Eq.~\eqref{eq:sum-w-sq} is $1/N$, and Eq.~\eqref{eq:alpha-ci-3} is divided by $N$.
  \item It tends to enhance the vector $\Im[\alpha_1^{6p}(\omega)]$ and tensor contributions $\Im[\alpha_2^{6p}(\omega)]$, because for arbitrary weights (different from 0, 1, and $1/N$), the three $J''$-terms of Eq.~\eqref{eq:alpha-ci-2} do not exactly compensate each other. 
\end{itemize}

The weights $w_{\beta''m}^{(J'')}$ associated with the eigenvectors of excited energy levels are therefore crucial to calculate the imaginary part of the polarizability. In our previous work on erbium \cite{lepers2014}, we described the odd-parity levels with the configurations $4f^{12}6s6p$, $4f^{11}5d6s^2$ and $4f^{11}5d^26s$, which is likely to yield a reliable calculation of the weights $w_{\beta'',1}^{(J'')}$, that play an important part in the polarizability. By contrast, we did not consider the configurations $4f^{q-1}5d^26s$ for Dy ($q=10$) and Ho ($q=11$), because of the large number of levels belonging to those configurations. Since some of the weights $w_{\beta''m}^{(J'')}$ are not correct, our computed imaginary polarizabilities must be taken with caution. The relatively small ratio of the vector and tensor contributions with respect to the scalar one, observed in Ref.~\cite{li2017} and subsection \ref{sub:pola}, may be due to the lack of CI in our eigenvectors. In the next section, we will present a method to estimate the weights $w_{\beta'',1}^{(J'')}$ from experimental values of transition probabilities.

\subsection{Estimate of configuration-interaction mixing} 
\label{sub:ci-calc}

We consider transition probabilities $A_{\beta''J'',J0}$, characterizing the spontaneous emission from the level $|\beta''J''\rangle$ towards the ground level $|0J\rangle$, which are given by Eq.~\eqref{eq:gamma-nat} with $|\tilde{\beta}\tilde{J}\rangle = |0J\rangle$. Assuming that the transition is due to the coupling between the basis states $|1J''\rangle$ and $|0J\rangle$, we obtain that the squared transition dipole moment is proportional to $w_{\beta'',1}^{(J'')}$, see Eq.~\eqref{eq:dipsq-gs}, and so the Einstein coefficient is proportional to $A_{\beta''J'',0J} \propto w_{\beta'',1}^{(J'')} \times (E_{\beta''J''}-E_{0J})^3$. Supposing all transition energies approximately equal, \textit{i.e.}~$E_{\beta''J''}-E_{0J} \approx \hbar\omega_{6p}$, yields that the sum of transition probabilities for given $J$ and $J''$ is a $J$- and $J''$-independent constant,
\begin{equation}
  \sum_{\beta''} A_{\beta''J'',0J} \approx
   \frac{2\omega_{6p}^3 e^2r_{6s,6p}^2} {3\pi\epsilon_0\hbar c^3}
  \,.
  \label{eq:sum-aik}
\end{equation}
Therefore, knowing the transition energies and transition probabilities, we can express the weight $w_{\beta'',1}^{(J'')}$ as 
\begin{equation}
  w_{\beta'',1}^{(J'')} = \frac{ \dfrac{A_{\beta''J'',0J}}
   {(E_{\beta''J''}-E_{0J})^3}}
  {\sum_{\beta^*} \dfrac{A_{\beta^*J'',0J}}
   {(E_{\beta^*J''}-E_{0J})^3}} \,
  \label{eq:w-exp}
\end{equation}
where the terms $(E_{\beta''J''}-E_{0J})$ and $(E_{\beta^*J''}-E_{0J})$ have been explicitly written, in order to get a better estimate of $w_{\beta'',1}^{(J'')}$, even though they could be approximated by $\hbar\omega_{6p}$.

In practice, J.~Lawler and E.~Den Artog's group performed extensive measurements of transition probabilities, especially in dysprosium \cite{wickliffe2000}, holmium \cite{nave2003}, erbium \cite{lawler2010} and thulium \cite{wickliffe1997}. The spectrum of the ground level is composed of a forest of weak transitions from which emerge a few strong transitions with similar transition energies. The number of strong lines (say with $A_{\beta''J'',0J} > 10^7$ s$^{-1}$) increases with increasing atomic number. When calculating the sum of Einstein coefficients for separated $J''$, see Eq.~\eqref{eq:sum-aik}, one usually finds 2.1 to $2.4 \times 10^8$ s$^{-1}$. Among those transitions, some are certainly not due to $6s$-$6p$, but rather to $4f$-$5d$ excitation; however they are so weak that they will not affect the calculation of $w_{\beta'',1}^{(J'')}$ with Eq.~\eqref{eq:w-exp}.

\begin{table}
  \caption{Comparison of theoretical and experimental energies of some selected excited odd-parity levels $|\beta''J''\rangle$ of erbium, of transition probabilities characterizing the spontaneous emission from the levels $|\beta''J''\rangle$ to the ground level $|0J\rangle = |4f^{12}6s^2 \;^3H_6\rangle$, and of the weight of the component $|1J''\rangle = |4f^{12}(^3H_6) .6s6p(^1P_1^o) \,(6,1)^o_{J''}\rangle$ in the eigenvector associated with the level $|\beta''J''\rangle$, see Eq.~\eqref{eq:ci-wf}. The theoretical quantities, in the columns ``Th.'', come from our previous work \cite{lepers2014}, whereas the experimental ones, in the column ``Exp.'', come from Ref.~\cite{lawler2010}. The experimental weights $w_{\beta'',1}^{(J'')}$ are given by Eq.~\eqref{eq:w-exp}. The selected excited levels are such that the experimental transition probability towards the ground level is larger than $10^7$ s$^{-1}$. The notation ($n$) stands for $\times 10^n$.}
  \label{tab:weight-er}
  \begin{ruledtabular}
  \begin{tabular}{rrcrrrr}
  \multicolumn{2}{c}{$E_{\beta''J''}$ (cm$^{-1}$)} & \multirow{2}{*}{$\quad J'' \quad$} & \multicolumn{2}{c}{$A_{\beta''J'',0J}$ (s$^{-1}$)} & \multicolumn{2}{c}{$w_{\beta'',1}^{(J'')}$ (\%)} \\
  \cline{1-2} \cline{4-5} \cline{6-7}
  Exp. & \phantom{$_A^{^A}$}Th. & & \phantom{aaa} Exp. & \phantom{aaaa} Th. & \phantom{a} Exp. & \phantom{aa} Th. \\
  \hline                                                                   
   24083 & 24056 & 5 & 1.02(8) & 9.34(7) & 48 & 46 \\
   24457 & 24492 & 6 & 3.26(7) & 2.16(7) & 16 & 11 \\
   24943 & 24946 & 7 & 1.85(8) & 2.08(8) & 76 & 79 \\
   25159 & 25168 & 7 & 4.03(7) & 1.27(7) & 16 &  5 \\
   25163 & 25171 & 5 & 3.76(7) & 4.60(7) & 15 & 16 \\
   25393 & 25419 & 6 & 3.19(7) & 1.86(7) & 14 &  7 \\
   25598 & 25570 & 7 & 1.51(7) & 5.50(6) &  6 &  2 \\
   25682 & 25598 & 5 & 6.3(7)  & 4.28(7) & 24 & 13 \\
   25880 & 26071 & 6 & 1.22(8) & 9.68(7) & 49 & 31 \\
   26237 & 26178 & 6 & 2.90(7) & 8.43(7) & 11 & 26 \\
  \end{tabular}
  \end{ruledtabular}
\end{table}

In the case of erbium, we modeled the erbium spectrum including the configurations $4f^{12}6s6p$, $4f^{11}5d6s^2$ and $4f^{11}5d^26s$ \cite{lepers2014}, while we did not include either $4f^95d^26s$ for dysprosium \cite{li2017}, or $4f^{10}5d^26s$ for holmium (see section \ref{sec:ho-spec}). So for erbium, the ``experimental'' weights, given by Eq.~\eqref{eq:w-exp}, can be compared with the ``theoretical'' ones, that we can extract from our modeling of the spectrum \cite{lepers2014}. The results are given in Table \ref{tab:weight-er} for the odd-parity levels giving the strongest transitions (with probabilities larger than $10^7$\;s$^{-1}$) towards the ground level $4f^{12}6s^2 \;^3H_6$. In Table \ref{tab:weight-er} we also compare energies and transition probabilities. As discussed in Ref.~\cite{lepers2014}, the agreement on energy is very good. As for transition probabilities, the overall agreement is satisfactory, even if the theoretical transition probabilities and weights are globally smaller than the experimental ones. For a given level $|\beta''J''\rangle$, the discrepancies on $A_{\beta''J'',0J}$ and for $w_{\beta'',1}^{(J'')}$ are actually similar. This confirms our assumption that the strongest transitions are due to the $|1J''\rangle \to |0J\rangle$ components. This also means that, taking the experimental transition probabilities as benchmarks, we may improve our theoretical values by improving the quality of our eigenvectors.

\begin{table*}
  \caption{Dynamic dipole polarizabilities of dysprosium, holmium and erbium in their ground level, at the commonly used 1064-nm trapping wavelength. Namely we give the real part of the scalar contribution (in atomic units, $1\, \mathrm{a.u.} = e^2a_0^3 / 4\pi\epsilon_0$), as well as the imaginary part of the scalar, vector and tensor contributions (in $10^{-7}$\,a.u.). The columns ``Th.'' and ``Exp.'' stand for theoretical (see Refs.~\cite{li2017, lepers2014} and subsection \ref{sub:pola}) and experimental \cite{wickliffe2000, nave2003, lawler2010} transition energies and transition probabilities respectively. The columns ``Eq.\,\eqref{eq:w-exp}'' correspond to the application of Eqs.~\eqref{eq:alpha-k-pp-1}, \eqref{eq:alpha-k-pp-2} and \eqref{eq:w-exp}.}
  \label{tab:pola-dyhoer}
  \begin{ruledtabular}
  \begin{tabular}{ccrrrrrrrrr}
  \multirow{2}{*}{\; part \;} & \multirow{2}{*}{\; contrib. \;} & \multicolumn{3}{c}{Dy ($^5I_8$)} & \multicolumn{3}{c}{Ho ($^4I_{15/2}^o$)} & \multicolumn{3}{c}{Er ($^3H_6$)} \\
  \cline{3-5} \cline{6-8} \cline{9-11}
  & & \phantom{(\;\,$^{^A}_A$33)}Th. & \phantom{(\;\,3)}Exp. & \;Eq.\,\eqref{eq:w-exp} & \phantom{(\;\,33)}Th. & \phantom{(\;\,3)}Exp. & \;Eq.\,\eqref{eq:w-exp} & \phantom{(\;\,33)}Th. & \phantom{(\;\,3)}Exp. & \;Eq.\,\eqref{eq:w-exp} \\
  \hline                                                                   
   real & scalar & 193 & 177 & 188 & 187 & 160 & 186 & 164 & 155 & 170 \\
   imag. & scalar & 49.1 & 40.3 & 48.8 & 39.6 & 34.7 & 46.6 & 23.4 & 22.0 & 27.1 \\
        & vector & 11.3 & 12.9 & 15.2 & 19.1 & 17.0 & 17.1 & 17.4 & 11.2 & 12.4 \\
        & tensor & 5.8 & -9.0 & -11.3 & 4.9 & 5.5 & 9.2 & -6.9 & -5.4 & -5.0 \\
  \end{tabular}
  \end{ruledtabular}
\end{table*}

To illustrate the validity of our weight calculations, in Table \ref{tab:pola-dyhoer}, we give the real part of the scalar contribution, as well as the imaginary part of the scalar, vector and tensor contributions of the dynamic dipole polarizability at the frequency corresponding to a 1064-nm wavelength, for erbium, holmium and dysprosium. The calculations are carried out using three different methods. (i) The transition energies and squares of the transition dipole moments are taken from our full numerical modeling of the atomic spectra. In particular, the squares of the transition dipole moments are extracted from Einstein coefficients, by reversing Eq.~\eqref{eq:gamma-nat},
\begin{equation}
  \dipsq{\beta''J''}{0J} = \frac{3\pi\epsilon_0 \hbar^4 c^3
   (2J''+1) A_{\beta''J'',0J}}{(E_{\beta''J''}-E_{0J})^3} \,.
  \label{eq:dipsq-aik}
\end{equation}
This corresponds to the columns entitled ``Th.'' in Table \ref{tab:pola-dyhoer}. (ii) The transition energies and the squares of the transition dipole moments come from experimental measurements of transition probabilities using Eq.~\eqref{eq:dipsq-aik}; this corresponds to the columns entitled ``Exp.'' in Table \ref{tab:pola-dyhoer}. (iii) Polarizabilities are calculated using Eqs.~\eqref{eq:alpha-k-pp-1} and \eqref{eq:alpha-k-pp-2}; to that end, the weights $w_{\beta'',1}^{(J'')}$ are calculated by applying Eq.~\eqref{eq:w-exp} with experimental data, and the quantities $r_{6s6p}$ come from our fitting procedure of Einstein coefficients, namely $r_{6s6p}=3.551$ a.u.~for Er \cite{lepers2014}, 3.648 a.u.~for Dy \cite{li2017} and 3.630 a.u.~for Ho (see Sec.~\ref{sec:ho-spec}; for dipole moments, $1\, \mathrm{a.u.} = ea_0$, with $a_0$ the Bohr radius). This corresponds to the columns entitled ``Eq.~\eqref{eq:w-exp}'' in Table \ref{tab:pola-dyhoer}. The real part of the vector and tensor contributions are pointless here, as they vanish with method (iii).

First of all, we see that the real part of the scalar polarizability is smaller with the ``Exp.'' method. This is particularly striking in the case of holmium. In comparison with the ``Th'' method, this is due to the smaller number of experimental transitions than of theoretical ones. In contrast, the number of transitions in the ``Exp.'' and ``Eq.\,\eqref{eq:w-exp}'' methods is the same; however we saw in Table \ref{tab:weight-er} that the experimental weights are overestimated. Indeed there are certainly transitions with upper levels having a small $|1J''\rangle$ character which have not been detected. This result in the underestimation of the denominator of Eq.~\eqref{eq:w-exp}, and so the overestimation of $w_{\beta'',1}^{(J'')}$. Similar discrepancies are visible on the imaginary part of the scalar polarizability. Therefore it is appropriate to analyze the ratio vector-to-scalar and tensor-to-scalar contributions, in order to determine the anistropy of the photon-scattering rate.

Erbium is the atom for which this anisotropy is the most pronounced, for both the vector and the tensor contributions, even if the ratios vary significantly from one method to the other. From the method ``Eq.\,\eqref{eq:w-exp}'' to the method ``Th.'', the ratio $\Im(\alpha_\mathrm{vect}) / \Im(\alpha_\mathrm{scal})$ and $\Im(\alpha_\mathrm{tens}) / \Im(\alpha_\mathrm{scal})$ range from 0.46 and -0.18, to 0.74 and -0.29 respectively.

In the case of dysprosium, the agreement between the methods ``Exp'' and ``Eq.\,\eqref{eq:w-exp}'' is very good. The ratios $\Im(\alpha_\mathrm{vect}) / \Im(\alpha_\mathrm{scal})$ are equal to 0.32 and 0.31, and the ratios $\Im(\alpha_\mathrm{tens}) / \Im(\alpha_\mathrm{scal})$ to -0.22 and -0.23, respectively. With the ``Th.'' method, the ratios are smaller ($\Im(\alpha_\mathrm{vect}) / \Im(\alpha_\mathrm{scal}) = 0.23$ and $\Im(\alpha_\mathrm{tens}) / \Im(\alpha_\mathrm{scal}) = 0.12$), especially because this method does not allow for describing the CI-mixing in the levels at 23832 and 23878 cm$^{-1}$, and so it underestimates Eq~\eqref{eq:sum-w-sq}. 

Finally the case of holmium is hard to analyze, since no particular trend comes out of the calculations. The real part of the scalar polarizability is 27-a.u.~smaller in the ``Exp.'' method than in the two others. Moreover regarding the experimental transitions towards the ground level, none of them imply an upper level with an energy above 25571 cm$^{-1}$. These two facts suggests the possibility that some strong transitions have not been detected, especially with upper levels $J''=13/2$. For instance, in our full numerical modeling of Ho spectrum (see subsection \ref{sub:tp}) we predict two such transitions, with unobserved upper levels: one with $E^\mathrm{th}_{\beta'',13/2} = 28014$ cm$^{-1}$, $w_{\beta'',1}^{(13/2)} = 4$ \%, $A^\mathrm{th}_{\beta'',13/2,0J} = 2.61 \times 10^7$ s$^{-1}$, and the other with $E^\mathrm{th}_{\beta'',13/2} = 30942$ cm$^{-1}$, $w_{\beta'',1}^{(13/2)} = 7$ \%, $A^\mathrm{th}_{\beta'',13/2,0J} = 1.89 \times 10^7$ s$^{-1}$.

\section{Modeling of the holmium spectrum}
\label{sec:ho-spec}

In order to calculate the different components of the polarizabilities, and also the various $C_6$ coefficients, using the sum-over-state formulas, one needs an extensive set of transition energies and transition dipole moments. This section is devoted to the full numerical calculations of those quantities, in the case of holmium in its ground $^4I^o_{15/2}$ and first excited levels $^4I^o_{13/2}$. Indeed the transition between those two levels, allowed in the electric-quadrupole and magnetic-dipole approximations was suggested as a candidate for optical clocks \cite{kozlov2013, sukachev2016}, as those levels are expected to possess very similar polarizabilities.

As the principle of our calculations \cite{cowan1981, wyart2011, lepers2016} is identical to our previous work on dysprosium \cite{li2017}, we only highlight in this section the particularities of holmium. One of them is the rarity of experimental Land\'e g-factors, which gives to our work a predictive character in this respect. The experimental energies are published in the NIST database \cite{NIST_ASD}, constructed from the critical compilation of Martin \textit{et.~al.}~\cite{martin1978}, and from Ref.~\cite{kroger1997} which is posterior to the compilation. For odd-parity levels, we also use unpublished work from our group \cite{wyart2016}. Note that $^{165}$Ho, which is bosonic, possesses a nuclear spin $I=7/2$, but the resulting hyperfine structure is not considered in the present article.

\subsection{Energy levels}
\label{sub:ener-lev}

The ground level of holmium is of odd parity with the configuration $4f^{11}6s^{2}$, and total electronic angular momentum $J=15/2$. Table \ref{tab:oddlev} presents a comparison between our theoretical energies and Land\'e g-factors with their experimental counterparts. The theoretical values are obtained in a calculation including the configurations $4f^{11}6s^2$, $4f^{11}5d6s$ and $4f^{10}6s^26p$ \cite{wyart2016}. The levels of the $4f^{11}6s^2$ configuration can be labeled in the LS coupling scheme; for example, the orbital $L=6$ and spin $S=3/2$ angular momenta of the ground level are good quantum numbers up to 97 \%. By contrast, the level at 22593.53 cm$^{-1}$ is of $^4G$ and $^2H$ characters up to 44 and 36 \% respectively

\begin{table}
	\caption{Comparison of energies $E$ through the quantity $\Delta E = E^\mathrm{exp} - E^\mathrm{th}$ and Land\'e g-factors $g_L$ of Ho I odd-parity levels of the lowest electronic configuration [Xe]$4f^{11}6s^2$. The superscript {}``exp" stands for experimental values which are taken from \cite{NIST_ASD, kroger1997}. The superscript {}"th" stands for the theoretical values from the parametric study of Ref.~\cite{wyart2016}.}
	\label{tab:oddlev}
	\begin{ruledtabular}
	\begin{tabular}{rrrrrrr}
		\multirow{2}{*}{Term} & \multirow{2}{*}{$J$} & $E^\mathrm{exp}$ & $\Delta E$ & \multirow{2}{*}{$g_L^\mathrm{exp}$} & \multirow{2}{*}{$g_L^\mathrm{th}$} & \% leading \\
		 & & (cm$^{-1}$) & (cm$^{-1}$) & & & term \\
		\hline                                                                   
		\phantom{$^{^A}$}$^{4}I^o$ &  15/2 &        0        &           30 &          1.195 &             1.197 & 97 \\
		$^4I^o$ &  13/2 &     5419.70     &            7 &             -  &             1.107 & 99 \\ 
		$^4I^o$ &  11/2 &     8605.16     &           -6 &          1.012 &             0.985 & 85 \\
		$^4I^o$ &  9/2  &    10695.75     &           -5 &          0.866 &             0.864 & 60 \\
		$^4F^o$ &  9/2  &    13094.42     &           46 &             -  &             1.174 & 65 \\
		$^4G^o$ &  11/2 &    22593.53     &          -90 &             -  &             1.193 & 44 \\
	\end{tabular}
	\end{ruledtabular}
\end{table}

In the even parity, the electronic configurations included in our model are the two lowest ones $4f^{11}6s6p$ and $4f^{10}5d6s^{2}$ \cite{brewer1983}, which are connected to the ground-state configuration $4f^{11}6s^{2}$ by electric-dipole transitions. Therefore, in our model, we neglect the configuration interaction with other even-parity configurations, especially $4f^{10}5d^{2}6s$ which contains a large number of levels. By contrast, the first parametric study of even-parity levels was performed with configurations with a limited number of LS terms of the $4f^{10}$ and $4f^{11}$ cores, including configuration interaction with $4f^{10}5d^26s$; but such a truncation strongly damaged the quality of the Hamiltonian eigenvectors \cite{wyart1974}. In the present study, 92 even-parity levels were fitted to their known experimental counterparts \cite{NIST_ASD, kroger1997}, using 21 free energetic parameters, giving a 45-cm$^{-1}$ standard deviation. 

A comparison between theoretical and experimental levels is displayed in Table \ref{tab:evlev}, while the fitted parameters are given in Table \ref{tab:parev} (see appendix). Due to the lack of experimental g-factors data for most levels, we just list the theoretical results. All energies are given relative to the experimental $4f^{11}6s^{2}$ $^4I^o_{15/2}$ ground level. Despite the absence of the $4f^{10}5d^26s$ configuration, whose lowest classified level is at 20167.17 cm$^{-1}$, the agreement is very satisfactory.

\subsection{Transition probabilities}
\label{sub:tp}

Now that the energy parameters have been adjusted, the eigenvalues and eigenvectors of the Hamiltonian operator are fixed. The transition probabilities also depend on monoelectronic transition dipole moments (MTDMs) $-er_{n\ell,n''\ell''}$, whose adjustment using least-square fitting between theoretical and experimental transition probabilities, is the goal of this subsection.

Due to the configurations that we consider, there are two MTDMs coming into play: $r_{6s6p}$ and $r_{4f5d}$, corresponding respectively to the couples of configurations $4f^{11}6s^2$-$4f^{11}6s6p$ and $4f^{11}6s^2$-$4f^{10}5d6s^2$. The least-square fitting procedure between theoretical and experimental Einstein coefficients is performed on the scaling factors (SFs) $f_1 = r_{6s6p} / r_{6s6p}^\mathrm{HFR}$ and $f_2 = r_{4f5d} / r_{4f5d}^\mathrm{HFR}$, rather than the MTDMs themselves. This allows for more direct comparisons with results for dysprosium and erbium. Note that $r_{n\ell,n''\ell''}^\mathrm{HFR}$ stands for the \textit{ab initio} values calculated with Hartree-Fock method including relativistic corrections.

\begin{table}
  \caption{Transitions excluded from the least-square fitting procedure. The labels $|\beta''J''\rangle$ and $|\beta J\rangle$ correspond to upper and lower levels, respectively. The superscript {}``exp" stands for experimental values	which are taken from \cite{nave2003}. The transition wave number $\sigma_{\beta''J'',\beta J} = (E_{\beta''J''}-E_{\beta}) / 2\pi\hbar c$ is in the vacuum. The notation ($n$) stands for $\times 10^n$. A blank in the column {}``removal reason'' means that the upper level belongs neither to the $4f^{10}6s6p$ nor to the $4f^95d6s^2$ configuration, while {}``r.'' stands for {}``ratio''.}
	\label{tab:rml}                             
\begin{ruledtabular}
\begin{tabular}{rrrrrrr}
	$E_{\beta''J''}^\mathrm{exp}$ & \multirow{2}{*}{\quad $J''$} & $E_{\beta J}^\mathrm{exp}$ & \multirow{2}{*}{\quad $J$} & $\sigma_{\beta''J'',\beta J}^\mathrm{exp}$ & $A_{\beta''J'',\beta J}^\mathrm{exp}$ & \quad removal \\
	(cm$^{-1}$) & & (cm$^{-1}$) & & (cm$^{-1}$) & (s$^{-1}$) & reason \\
	\hline
	20258  & 6.5 &     0  & 7.5 &  20258 &  3.40(5) &      	\\
	24014  & 6.5 &     0  & 7.5 &  24014 &  1.06(8) & large r. \\
	24264  & 8.5 &     0  & 7.5 &  24264 &  1.42(7) & mixed$^\mathrm{a}$ \\
	24377  & 7.5 &     0  & 7.5 &  24377 &  5.78(6) & mixed$^\mathrm{b}$ \\
	24760  & 6.5 &     0  & 7.5 &  24760 &  1.20(6) &      	\\
	17059  & 6.5 &  5420  & 6.5 &  11640 &  0.34(3) & large r. \\
	18756  & 7.5 &  5420  & 6.5 &  13337 &  1.92(4) & small r. \\
	18858  & 6.5 &  5420  & 6.5 &  13438 &  0.91(4) & small r. \\
	20258  & 6.5 &  5420  & 6.5 &  14839 &  0.42(5) &     	\\
	24760  & 6.5 &  5420  & 6.5 &  19340 &  0.47(4) &      	\\
	25571  & 6.5 &  5420  & 6.5 &  20151 &  0.38(5) & small r. \\
	20241  & 6.5 &  8605  & 5.5 &  11636 &  1.19(4) & small r. \\
	20258  & 6.5 &  8605  & 5.5 &  11653 &  0.20(4) &      	\\
	22978  & 6.5 &  8605  & 5.5 &  14373 &  4.28(4) & small r. \\
	24760  & 6.5 &  8605  & 5.5 &  16155 &  0.48(4) &      	\\
\end{tabular}
\end{ruledtabular}
	$^\mathrm{a}$ mixed with level at 24361 cm$^{-1}$ \\
	$^\mathrm{b}$ mixed with level at 24661 cm$^{-1}$
\end{table}

As references, we take the measured transition probabilities of Ref.~\cite{nave2003}. We retain the transitions involving the ground and first excited levels, and upper levels with energies smaller than 30000 cm$^{-1}$. Indeed the levels above 30000 cm$^{-1}$ are hard to classify unambiguously in configurations $4f^{10}5d6s^2$ and $4f^{11}6s6p$. In addition, in the list of Ref.~\cite{nave2003}, we can see some strong transitions whose upper level does not belong to $4f^{11}6s6p$ or $4f^{10}5d6s^2$ configurations (according to the NIST database \cite{NIST_ASD}), \textit{e.g.}~$E_{\beta''J''}^\mathrm{exp}=24263.88$ cm$^{-1}$, $J'' = 17/2$, but is very close in energy to a $4f^{11}6s6p$ level with the same $J''$, \textit{e.g.}~$E_{\beta''J''}^\mathrm{exp} = 24360.81$ cm$^{-1}$. In contrast there is only one close theoretical level predicted, $E_{\beta''J''}^\mathrm{th} = 24354.1$ cm$^{-1}$. Similar to dysprosium, we can assume the eigenvector of that theoretical level contains some components of the $|1J''\rangle$ state which is shared by the two {}``real'' levels. In those particular cases, we compare our theoretical Einstein coefficient with the sum of experimental ones. In Table \ref{tab:rml}, the 2 transitions labeled {}``mixed'' correspond to that situation.

Due to strong differences between experimental and theoretical Einstein coefficients, we excluded 6 transitions (one with large ratio $A_{\beta''J'',\beta J}^\mathrm{th}/A_{\beta''J'',\beta J}^\mathrm{exp}$, while other four with very small ratios). Special attention should be paid to the strong transition between the ground level and the excited $J''=13/2$ level at $E_{\beta''J''}^\mathrm{exp} = 24014.2$ cm$^{-1}$. For optimal scaling factors $f_1$ and $f_2$ (see below), the error on the other strongest transitions (above $10^8$ s$^{-1}$) is below 5 \%, while it is 14 \% for the latter. This may be due to an underestimated experimental value. Another possible explanation is the following: there is a close $J''=6.5$ level, at $E_{\beta''J''}^\mathrm{exp} = 23955.69$ cm$^{-1}$; where comparing the sum of transition probabilities implying those two upper levels, the theory-experiment agreement is very good (1.47 and $1.42 \times 10^8$ s$^{-1}$ respectively). The agreement for individual transition can probably be improved by a better CI-mixing between the configurations $4f^{11}6s6p$ and $4f^{10}5d6s^2$.

\begin{table}
	\caption{Comparison of Einstein-$A$ coefficients.
		The superscript {}``exp" stands for experimental values
		which are taken from \cite{nave2003}. The superscript {}``th'' 
		stands for the theoretical values from the present 
		calculations. The notation ($n$) stands for $\times 10^n$. Values with an asterisk ($^*$) correspond to sums of experimental Einstein coefficients (see Table \ref{tab:rml}).}
	\label{tab:aik}                           
\begin{ruledtabular}
\begin{tabular}{rrrrrrr}
	$E_{\beta''J''}^\mathrm{exp}$ & \multirow{2}{*}{\quad $J''$} & $E_{\beta J}^\mathrm{exp}$ & \multirow{2}{*}{\quad $J$} & \quad $\sigma_{\beta''J'',\beta J}^\mathrm{exp}$ & $A_{\beta''J'',\beta J}^\mathrm{exp}$ & $A_{\beta''J'',\beta J}^\mathrm{th}$ \\
	(cm$^{-1}$) & & (cm$^{-1}$) & & (cm$^{-1}$) & (s$^{-1}$) & (s$^{-1}$) \\
	\hline 
	16710  & 8.5 &     0  & 7.5 &  16710 & 	 9.20(5) &  1.51(6)	\\
	16882  & 7.5 &     0  & 7.5 &  16882 &	 3.60(5) &  7.91(5)	\\
	17059  & 6.5 &     0  & 7.5 &  17059 &	 6.50(5) &  1.25(6)	\\
	18652  & 7.5 &     0  & 7.5 &  18652 &	 2.99(5) &  2.19(5)	\\
	18756  & 7.5 &     0  & 7.5 &  18756 &	 2.20(5) &  5.77(4)	\\
	18858  & 6.5 &     0  & 7.5 &  18858 &	 2.70(5) &  2.87(5)	\\
	20075  & 7.5 &     0  & 7.5 &  20075 &	 1.11(6) &  6.24(5)	\\
	20241  & 6.5 &     0  & 7.5 &  20241 &	 2.15(6) &  1.72(6)	\\
	22978  & 6.5 &     0  & 7.5 &  22978 &	 9.30(6) &  1.18(7)	\\
	23445  & 7.5 &     0  & 7.5 &  23445 &	 3.70(6) &  3.62(6)	\\
	23499  & 8.5 &     0  & 7.5 &  23499 &	 1.00(7) &  1.13(7)	\\
	23835  & 7.5 &     0  & 7.5 &  23835 &	 3.88(6) &  2.79(6)	\\
	23956  & 6.5 &     0  & 7.5 &  23956 &	 3.12(7) &  2.23(7)	\\
	24361  & 8.5 &     0  & 7.5 &  24361 &	 2.18(8)$^*$ &  2.18(8)	\\
	24661  & 7.5 &     0  & 7.5 &  24661 &	 2.06(8)$^*$ &  2.14(8)	\\
	24741  & 6.5 &     0  & 7.5 &  24741 &	 4.48(7) &  3.08(7)	\\
	25273  & 7.5 &     0  & 7.5 &  25273 &	 6.30(6) &  8.32(6)	\\
	25571  & 6.5 &     0  & 7.5 &  25571 &	 5.24(5) &  5.08(5)	\\
	16882  & 7.5 &  5420  & 6.5 &  11463 &	 6.00(3) &  4.92(3)	\\
	20075  & 7.5 &  5420  & 6.5 &  14655 &	 5.40(4) &  7.48(3)	\\
	20241  & 6.5 &  5420  & 6.5 &  14822 &	 2.58(5) &  3.38(4)	\\
	22413  & 5.5 &  5420  & 6.5 &  16993 &	 8.70(5) &  1.51(6)	\\
	22978  & 6.5 &  5420  & 6.5 &  17558 &	 2.60(5) &  3.94(4)	\\
	24741  & 6.5 &  5420  & 6.5 &  19321 &	 3.52(5) &  8.96(4)	\\
	25273  & 7.5 &  5420  & 6.5 &  19853 &	 2.99(4) &  1.82(4)	\\
	29070  & 5.5 &  5420  & 6.5 &  23650 &	 1.06(8) &  1.11(8)	\\
	29097  & 5.5 &  5420  & 6.5 &  23677 &	 6.70(6) &  2.80(6)	\\
	29643  & 7.5 &  5420  & 6.5 &  24223 &	 2.12(8) &  2.13(8)	\\
	29752  & 6.5 &  5420  & 6.5 &  24332 & 	 2.00(8) &  1.92(8)	\\
	25571  & 6.5 &  8605  & 5.5 &  16966 &	 7.50(5) &  1.39(6)	\\
\end{tabular}
\end{ruledtabular}
\end{table}

We fitted the SFs using the remaining 29 transitions, and found optimal scaling factors $f_1=0.798$, $f_2=0.969$, corresponding to a standard deviation on Einstein coefficients (see Ref.~\cite{lepers2014}, Eq.~(15)) $\sigma_A = 4.14\times 10^6$ s$^{-1}$. In particular the 5 strongest transitions are calculated with a precision better than 5 \%. Then, because the experimental Einstein coefficients in Ref.~\cite{nave2003} are given with uncertainties reaching up to 10 \%, we made 10000 fits  in which all the experimental $A$-coefficients have a random value within their uncertainty range. We obtain optimal scaling factors with statistical uncertainties: $f_1 = 0.799 \pm 0.010$ and $f_2 = 0.97 \pm 0.24$. The standard deviation is therefore much more sensitive to $r_{6s6p}$ than to $r_{4f5d}$, since it involves the strongest transitions \cite{lepers2014, lepers2016, li2017}. In what follows, we take the optimal scaling factors $f_1=0.799$ and $f_2=0.97$, for which a comparison between experimental and theoretical transition probabilities involving the two lowest levels of Ho I are presented in Table \ref{tab:aik}. Using those optimal SFs, we can also calculate transition probabilities which have not been measured, and which are available upon request to the authors. In particular, as discussed in subsection \ref{sub:ci-calc}, we predict two strong transitions with unobserved upper levels of $J''=13/2$.

\subsection{Dynamic dipole polarizability}
\label{sub:pola}

The optimal set of spectroscopic data obtained in the previous subsection will now be used to compute the real and imaginary parts of the scalar, vector and tensor polarizabilities given by Eqs.~\eqref{eq:alpha-scal-re}--\eqref{eq:alpha-tens-im}. The squared transition dipole moments appearing in the sum are extracted from theoretical Einstein coefficients using Eq.~\eqref{eq:dipsq-aik}.

To compare our results with literature, the scalar, vector and tensor static dipole polarizabilities are presented in Table \ref{tab:pola-re-im-freq}, as well as the dynamic ones for the widespread laser-trapping wavelength $\lambda=1064$ nm (corresponding to a wave number $\sigma=9398$ cm$^{-1}$). As one can notice for the ground-level scalar polarizability, agreement is good between the different theoretical results and with the new experimental one (for which we do not have any numerical value \cite{ma2015}). The tensor static polarizability is much smaller than the scalar one in all sources. For the $^4I^o_{13/2}$ level there are no literature values to our knowledge. They are actually very similar to those of the ground level, which supports the possibility to use those levels in a clock transition.

\begin{table*}
	\caption{Real and imaginary parts of the scalar, vector and tensor dynamic dipole polarizabilities, at wave numbers $\sigma = \omega/2\pi c = 0$ and 9398 cm$^{-1}$ (corresponding to $\lambda=1064$ nm), for the ground $^4I^o_{15/2}$ and first excited level $^4I^o_{13/2}$ of holmium. Our results are compared with available literature values.
\label{tab:pola-re-im-freq}}
\centering
\begin{ruledtabular}
\begin{tabular}{cccccccc}
  \multirow{2}{*}{level} & $\sigma$  
   & \multicolumn{3}{c}{Real part (a.u.)} 
   & \multicolumn{3}{c}{Imaginary part ($10^{-7}$~a.u.)} \\
  \cline{3-8}
   & (cm$^{-1})$ & scalar & vector & tensor 
   & scalar & vector & tensor \\
  \hline
  \phantom{$^{^A}$}$^4I^o_{15/2}$ \phantom{$^{^A}$} & 0 & 160 & 0 & -2.3 & 25.1 & 0 & 3.4 \\
   & & 159 \cite{lide2012}, 170 \cite{chu2007} & 
   & -3.19 \cite{chu2007} & & & \\
   & & 156 \cite{dzuba2014}, 161 \cite{dzuba2016} & 
   & -1.17 \cite{rinkleff1994} & & & \\
   & 9398 & 187 & 1.1 & -3.5 & 39.6 & 19.1 & 4.9 \\
  \hline
  \phantom{$^{^A}$} $^4I^o_{13/2}$ \phantom{$^{^A}$} & 0 & 160 & 0 & -2.0 & 24.1 & 0 & 1.4 \\
   & 9398 & 187 & 1.0 & -3.0 & 38.3 & 17.7 & 2.0 \\
\end{tabular}
\end{ruledtabular}
\end{table*}

For both levels, the main result obtained in our previous work on erbium \cite{lepers2014} and dysprosium \cite{li2017} is confirmed. Regarding the real part, the vector and tensor polarizability are much smaller than the scalar one. The trapping potential is thus mostly isotropic, as it hardly depends on the electric-field polarization or the atomic azimuthal quantum number. By contrast, the tensor, and especially vector contributions of the imaginary part represent a significant fraction of the scalar contribution, which makes photon-scattering anisotropic. In subsection \ref{sub:ci-calc}, this anisotropy is discussed in details and compared with the results on neighboring atoms.

\begin{figure}[h!]
  \includegraphics[width=7cm]{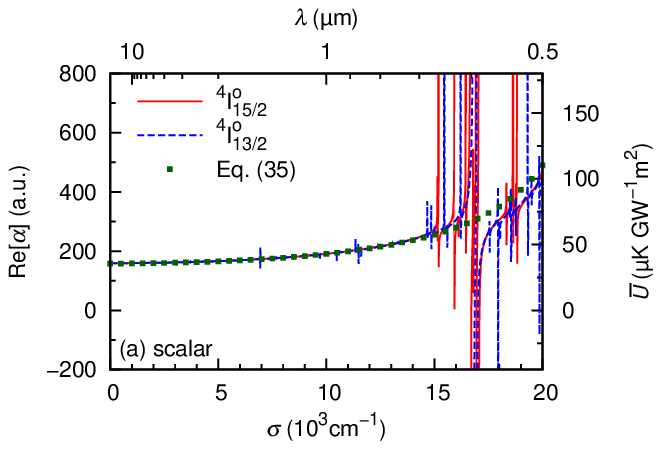} 
  \includegraphics[width=7cm]{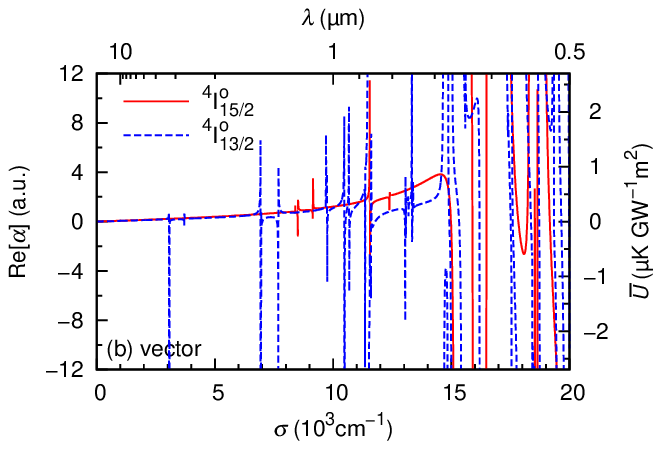}
  \includegraphics[width=7cm]{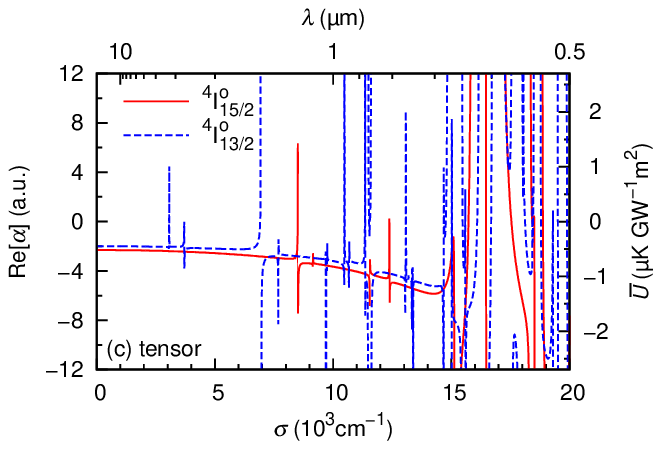}
  \caption{(Color online) Real part of the (a) scalar, (b) vector and (c) tensor dynamic dipole polarizabilities of the $^4I^o_{15/2}$ and $^4I^o_{13/2}$ levels in atomic units and corresponding trap depths obtained for an intensity of 1 GW.m$^{-2}$, as functions of the trapping wave number $\sigma$ and wavelength $\lambda$. Panel (a) also displays the real part of the scalar polarizability given by Eq.~\eqref{eq:alpha-ci-1}.
  \label{fig:pola-real-freq}}
\end{figure}

\begin{figure}[h!]
  \includegraphics[width=7cm]{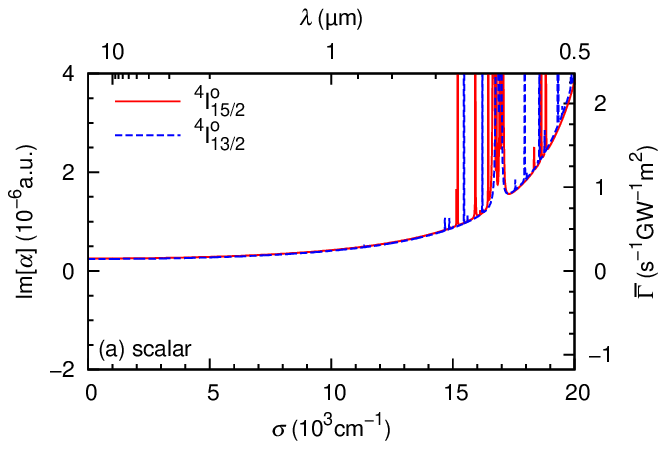} 
  \includegraphics[width=7cm]{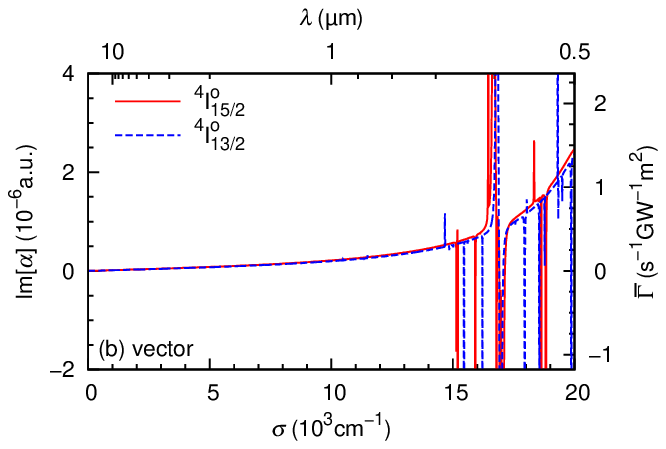}
  \includegraphics[width=7cm]{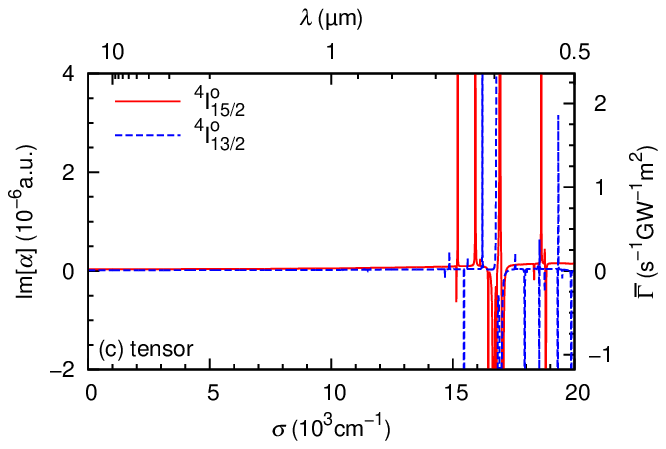}
	\caption{(Color online) Imaginary part of the (a) scalar, (b) vector and (c) tensor dynamic dipole polarizabilities of the $^4I^o_{15/2}$ and $^4I^o_{15/2}$ levels in atomic units and corresponding photon-scattering rates obtained for an intensity of 1 GW.m$^{-2}$, as functions of the trapping wave number $\sigma$ and wavelength $\lambda$.
	\label{fig:pola-img-freq}}
\end{figure}

Those features are confirmed on Figures \ref{fig:pola-real-freq} and \ref{fig:pola-img-freq}, which present the real, resp.~imaginary, parts of the scalar, vector and tensor polarizabilities as functions of the field wavelength $\lambda$ and wave number $\sigma = 1/\lambda = \omega/2\pi c$ ($c$ being the speed of light). We present our results in atomic units and in the corresponding relevant quantities in physical units. The real part of the polarizability is associated with the potential energy $\overline{U}$, in equivalent temperatures of microkelvins ($\mu$K), obtained for a laser intensity of 1~GW/m$^2$. The imaginary part is associated with the photon-scattering rate $\overline{\Gamma}$, in inverse seconds (s$^{-1}$), for the same intensity. 
In Fig.~\ref{fig:pola-real-freq} (a), we also compare the real part of the scalar DDP given by our full numerical results and by the simplified model of Eq.~\eqref{eq:alpha-ci-1} with $\omega_{6p} = 24320$ cm$^{-1}$ and $r_{6s6p} = 3.630$ a.u.. We see that the latter reproduces very nicely the background polarizability for both levels, but not the narrow peaks due to transition toward the levels of the $4f^{10}5d6s^2$ configuration, or to the intercombination lines toward the levels $4f^{11}(^4I^o_{15/2}) . 6s6p(^3P^o_1) \, (15/2,1)_{J''}$ around 17000 cm$^{-1}$.

\subsection{Van der Waals $C_6$ coefficients}
\label{sub:vdw}

Using the optimal spectroscopic data of subsection \ref{sub:tp}, we can also compute the van der Waals $C_6$ coefficients, which also consist of a sum on transition energies and transition dipole moments \cite{stone1996, kaplan2006}. The van der Waals interaction between two open-shell atoms actually depends on a limited numbers of coefficients $C_{6,k_Ak_B}$, where the indexes $k_A$ and $k_B$ correspond to the rank of irreducible tensors \cite{li2017, lepers2017}. Giving diagonal matrix elements, the coefficient $C_{6,00}$ is referred to as isotropic, while the other $C_{6,k_Ak_B}$ are called anisotropic. Table \ref{tab:c6} displays our calculated $C_{6,k_Ak_B}$ coefficients, for the two lowest levels of holmium. For any pair of levels, and similarly to the case of erbium \cite{lepers2014} and dysprosium \cite{li2017}, the isotropic coefficient $C_{6,00}$ strongly dominates the anisotropic ones. Moreover, the coefficients are very similar for the three pairs of levels, because the polarizabilities of the levels $^4I^o_{15/2}$ and $^4I^o_{13/2}$ are almost equal (see Table \ref{tab:pola-re-im-freq}).

\begin{table}
  \caption{$C_6$ coefficients in atomic units, characterizing the van der Waals interactions between holmium atoms in the ground $^4I^o_{15/2}$ or first excited level $^4I^o_{13/2}$, as functions of the pairs of indexes $k_A$, $k_B$ \cite{li2017, lepers2017}. The case $k_A=k_B=0$ corresponds to the so-called isotropic $C_6$ coefficient \cite{stone1996}. The {}``0'' corresponds to an absolute value smaller than 0.1 a.u..
  \label{tab:c6}}
  \centering
  \begin{ruledtabular}
  \begin{tabular}{cccc}
     $k_A$, $k_B$ & $^4I^o_{15/2}$ -- $^4I^o_{15/2}$ & $^4I^o_{15/2}$ -- $^4I^o_{13/2}$
      & $^4I^o_{13/2}$ -- $^4I^o_{13/2}$ \\
    \hline
     0, 0 & -2214 & -2214 & -2214  \\
     1, 1 &     0 &     0 &     0  \\
     2, 0 &  -8.0 &  -8.0 &  -7.3  \\
     0, 2 &  -8.0 &  -7.3 &  -7.3  \\
     2, 2 &     0 &     0 &     0  \\
  \end{tabular}
  \end{ruledtabular}
\end{table}

\section{Conclusion}
\label{sec:conclu}

In this article, we have derived a simplified model to characterize the optical trapping of ultracold lanthanide atoms. We have calculated analytically the real and imaginary parts of the scalar, vector and tensor polarizabilities, assuming that the transitions involving valence electrons are the only ones present in the sum-over-state formula. We have given an analytical expression for the contribution from all the levels belonging to a given electronic configuration; that expression only depends on two parameters: an effective transition energy and an effective transition dipole moment.
When applied to the two lowest levels of holmium, our model nicely reproduces the calculations based on the detailed modeling on the even-parity levels of holmium. We expect our simplified model to properly estimate the polarizabilities of levels of the $4f^q6s6p$ configuration, which are relevant for trapping, but which involve highly excited levels, for example of configurations $4f^q6s6d$ and $4f^q6p^2$, rarely known experimentally.
In this respect, the future study on thulium will be particularly interesting, since high-lying excited states have been characterized in details, and so the simplified expressions given in the present paper will be compared with those involving individual transition energies and transition dipole moments.

Regarding the ground level of lanthanides, we have also studied the influence of configuration interaction between $4f^q6s6p$ and other configurations. We have demonstrated that the real part of the polarizability is insensitive to configuration interaction, and that the vector and tensor polarizabilities are vanishingly small. By contrast, the imaginary part turns out to be very sensitive to configuration interaction among excited levels; the latter is responsible for a decrease in the scalar contribution, and for an increase in the vector and tensor ones. By comparing our numerical results for dysprosium, holmium and erbium, we have found significant variations of the imaginary part of the various polarizabilities. For example the scalar contribution for erbium is roughly twice as small as for dysprosium, which tends to prove that configuration-interaction mixing is stronger in erbium. This is very surprising, as the density of levels of dysprosium is larger around 25000 cm$^{-1}$, and so \textit{a priori} more favorable to configuration interaction. Again, the case of thulium will be particularly enlightening, because the large number of relatively strong transitions (with probabilities larger than $10^7$ s$^{-1}$, see Ref.~\cite{wickliffe1997}) suggests even stronger configuration interaction than in erbium. Finally we would like to highlight that experimental measurements of heating rate or trap lifetimes are particularly welcome, in order to check the validity of our predictions.

\section*{Acknowledgements}

The authors acknowledge support from {}``DIM Nano-K'' under the project {}``InterDy'', and from {}``Agence Nationale de la Recherche'' (ANR) under the project {}``COPOMOL'' (contract ANR-13-IS04-0004-01).

\appendix

\section{Even-parity nergy levels}

This appendix presents the detailed calculations of holmium even-parity levels. Table \ref{tab:evlev} presents the results of our calculations, including to the discrepancy between theoretical and experimental energies. Table \ref{tab:parev} contains the optimal parameters after the least-square fitting procedure on energies.

\begin{longtable*}{rrrllr}
	\caption{Same as Table \ref{tab:oddlev} for Ho I even-parity levels. The theoretical values $E^\mathrm{th}$, the Land\'e g-factors	$g_L^\mathrm{th}$ and the percentage of configurations and LS terms are derived by means of the Cowan code called ``RCG'' with the parameter set reported in Table \ref{tab:parev}. In the configuration notations, $A$ stands for $4f^{11}$, $B$ for $4f^{10}$, $ds^2$ for $5d6s^2$, $sp$ for $6s6p$. The lower-case letters or Arabic numbers appearing in the seventh column correspond to different possible parent terms \cite{cowan1981}. The terms in parentheses are associated with the core configurations $A$ or $B$.}
	\label{tab:evlev}                             
	\endfirsthead
	\caption{even parity levels of Ho I (continued)} \\
	\hline
	$E^\mathrm{exp}$ & $E^\mathrm{th}$ & $\Delta E$ & \multirow{2}{*}{$g_L^\mathrm{th}$} & Leading & \% leading   \\
	(cm$^{-1}$) & (cm$^{-1}$) & (cm$^{-1}$) &  & configuration & LS term \\
	\hline
	\endhead
	\hline
	$E^\mathrm{exp}$ & $E^\mathrm{th}$ & $\Delta E$ & \multirow{2}{*}{$g_L^\mathrm{th}$} & Leading & \% leading   \\
	(cm$^{-1}$) & (cm$^{-1}$) & (cm$^{-1}$) &  & configuration & LS term \\
	\hline
	& & & $J=7/2$ & &            \\
	32758.37  &  32750.2    &    8   & 1.253 & $B-ds^2$  &   15 $B-ds^2 \, (^5S) ^4D$   \\
	32931.51  &  32935.7    &   -4   & 1.232 & $B-ds^2$  &   25 $B-ds^2 \, (^5F) ^4G$   \\
	33188.42  &  33202.1    &  -14   & 0.926 & $B-ds^2$  &   37 $B-ds^2 \, (^5F) ^4H$   \\
	35078.45  &  35105.5    &  -27   & 1.147 & $B-ds^2$  &   19 $B-ds^2 \, (^5G) ^6D$   \\
	36001.87  &  35991.9    &   10   & 1.074 & $B-ds^2$  &   13 $B-ds^2 \, (^5G) ^6G$   \\
	36504.15  &  36486.6    &   18   & 1.114 & $B-ds^2$  &    9 $B-ds^2 \, (^5G) ^4F$   \\  
	\hline 
	&& & $J=9/2$ & &      \\
	16719.62  &  16727.3    &   -8    & 1.118 & $B-ds^2$  &   40 $B-ds^2 \, (^5I) ^6G$   \\
	18757.87  &  18682.1    &   76    & 0.991 & $B-ds^2$  &   47 $B-ds^2 \, (^5I) ^6I$   \\
	21373.01  &  21371.6    &    1    & 1.140 & $B-ds^2$  &   62 $B-ds^2 \, (^5I) ^4G$   \\
	23861.17  &  23909.6    &  -48    & 0.980 & $A-sp$    &   32 $A-sp   \, (^4I) ^6H$   \\
	24355.64  &  24367.3    &  -12    & 1.137 & $B-ds^2$  &   27 $B-ds^2 \, (^5I) ^4H$   \\
	25453.49  &  25427.1    &   26    & 0.906 & $A-sp$    &   30 $A-sp   \, (^4I) ^6I$   \\
	26039.99  &  25991.7    &   48    & 0.831 & $B-ds^2$  &   48 $B-ds^2 \, (^5I) ^4I$   \\
	28638.41  &  28653.1    &  -15    & 1.456 & $B-ds^2$  &   54 $B-ds^2 \, (^5S) ^6D$   \\
	32039.69  &  32053.7    &  -14    & 1.247 & $B-ds^2$  &   20 $B-ds^2 \, (^5G) ^6D$   \\
	33577.20  &  33553.1    &   24    & 1.159 & $B-ds^2$  &   18 $B-ds^2 \, (^5F) ^4G$   \\
	35270.88  &  35269.7    &    1    & 1.137 & $B-ds^2$  &   11 $B-ds^2 \, (^5G) ^6D$   \\ 
	\hline
	&  & & $J=11/2$ & &      \\  
	13082.93  &  13094.9    &  -12    & 1.260 & $B-ds^2$  &   50 $B-ds^2 \, (^5I) ^6G$   \\
	15792.13  &  15805.2    &  -13    & 1.143 & $B-ds^2$  &   37 $B-ds^2 \, (^5I) ^6I$   \\
	16937.43  &  16958.5    &  -21    & 1.244 & $B-ds^2$  &   52 $B-ds^2 \, (^5I) ^4G$   \\
	18491.21  &  18465.0    &   26    & 1.107 & $B-ds^2$  &   25 $B-ds^2 \, (^5I) ^6H$   \\
	18821.25  &  18802.1    &   19    & 1.107 & $A-sp$    &   46 $A-sp   \, (^4I) ^2H$   \\
	20493.40  &  20427.2    &   66    & 0.903 & $B-ds^2$  &   56 $B-ds^2 \, (^5I) ^6K$   \\
	20849.13  &  20863.0    &  -14    & 1.104 & $B-ds^2$  &   32 $B-ds^2 \, (^5I) ^4H$   \\
	22413.14  &  22378.1    &   35    & 1.078 & $A-sp$    &   29 $A-sp   \, (^4I) ^6H$   \\
	23379.31  &  23361.4    &   18    & 1.065 & $B-ds^2$  &   29 $B-ds^2 \, (^5I) ^4I$   \\
	23946.16  &  23961.3    &  -15    & 1.057 & $A-sp$    &   30 $A-sp   \, (^4I) ^6H$   \\
	24141.21  &  24179.4    &  -38    & 1.276 & $B-ds^2$  &   37 $B-ds^2 \, (^5F) ^6F$   \\
	24830.43  &  24907.2    &  -77    & 0.968 & $A-sp$    &   27 $A-sp   \, (^4I) ^6K$   \\
	25261.55  &  25271.4    &  -10    & 0.864 & $B-ds^2$  &   44 $B-ds^2 \, (^5I) ^4K$   \\
	25503.33  &  25467.7    &   36    & 0.998 & $A-sp$    &   14 $A-sp   \, (^4I) ^6K$   \\
	25914.31  &  25997.5    &  -83    & 1.209 & $B-ds^2$  &   32 $B-ds^2 \, (^5F) ^6H$   \\
	28793.03  &  28824.0    &  -31    & 1.230 & $B-ds^2$  &   48 $B-ds^2 \, (^5F) ^4G$   \\
	29069.78  &  29016.7    &   53    & 1.154 & $A-sp$    &   33 $A-sp   \, (^4I) ^4Hb$  \\
	29096.77  &  29132.9    &  -36    & 1.078 & $A-sp$    &   19 $A-sp   \, (^4F) ^6F$   \\
	30423.60  &  30332.7    &   91    & 1.240 & $B-ds^2$  &   39 $B-ds^2 \, (^5F) ^6G$   \\
	31903.28  &  31862.7    &   41    & 1.227 & $B-ds^2$  &   18 $B-ds^2 \, (^5F) ^6G$   \\
	32837.21  &  32860.6    &  -23    & 1.014 & $A-sp$    &   31 $A-sp   \, (^4I) ^4Ib$  \\
	33212.51  &  33287.1    &  -75    & 1.206 & $A-sp$    &   23 $A-sp   \, (^2H) ^4G2$  \\
	33986.71  &  33965.6    &   21    & 1.104 & $A-sp$    &   10 $A-sp   \, (^2H) ^2I2$  \\
	34270.67  &  34292.7    &  -22    & 1.141 & $B-ds^2$  &   19 $B-ds^2 \, (^5G) ^4G$   \\  
	\hline
	& & & $J=13/2$ & &      \\
     9147.08  &   9117.3    &   30    & 1.338 & $B-ds^2$  &   66 $B-ds^2 \, (^5I) ^6G$   \\
	12344.55  &  12364.8    &  -20    & 1.236 & $B-ds^2$  &   27 $B-ds^2 \, (^5I) ^6I$   \\
	15081.12  &  15112.6    &  -32    & 1.177 & $B-ds^2$  &   46 $B-ds^2 \, (^5I) ^4H$   \\
	16735.95  &  16763.6    &  -28    & 1.200 & $B-ds^2$  &   41 $B-ds^2 \, (^5I) ^6H$   \\
	17059.35  &  17019.9    &   39    & 1.193 & $A-sp$    &   29 $A-sp   \, (^4I) ^4Ha$  \\
	18564.90  &  18493.2    &   72    & 1.050 & $B-ds^2$  &   45 $B-ds^2 \, (^5I) ^6K$   \\
	18858.19  &  18792.4    &   66    & 1.138 & $A-sp$    &   37 $A-sp   \, (^4I) ^2I$   \\
	20241.31  &  20266.4    &  -25    & 1.104 & $B-ds^2$  &   43 $B-ds^2 \, (^5I) ^4I$   \\
	21044.81  &  21012.1    &   33    & 0.887 & $B-ds^2$  &   67 $B-ds^2 \, (^5I) ^6L$   \\
	21584.89  &  21623.2    &  -38    & 1.174 & $A-sp$    &   27 $A-sp   \, (^4I) ^6H$   \\
	22157.86  &  22214.2    &  -56    & 1.059 & $A-sp$    &   22 $A-sp   \, (^4I) ^4Ka$  \\
	22978.19  &  22969.7    &   08    & 1.013 & $B-ds^2$  &   43 $B-ds^2 \, (^5I) ^4K$   \\
	24014.22  &  23908.9    &  105    & 1.199 & $A-sp$    &   39 $A-sp   \, (^4I) ^4Hb$  \\
	23955.68  &  23968.7    &  -13    & 1.138 & $A-sp$    &   32 $A-sp   \, (^4I) ^6I$   \\
	24740.52  &  24739.9    &    1    & 1.245 & $B-ds^2$  &   36 $B-ds^2 \, (^5F) ^6H$   \\
	25571.15  &  25569.0    &    2    & 1.048 & $A-sp$    &   34 $A-sp   \, (^4I) ^6K$   \\
	25930.66  &  25950.9    &  -20    & 0.831 & $B-ds^2$  &   75 $B-ds^2 \, (^5I) ^4L$   \\
	27141.28  &  27143.8    &   -3    & 1.096 & $A-sp$    &   23 $A-sp   \, (^4I) ^6I$   \\
	29751.91  &  29792.1    &  -40    & 1.098 & $A-sp$    &   45 $A-sp   \, (^4I) ^4Ib$  \\
	33907.40  &  33780.7    &  127    & 1.196 & $A-sp$    &   22 $A-sp   \, (^2H) ^4H2$  \\       
	\hline      
	& & & $J=15/2$ & &      \\ 
	8427.11  &   8482.7     &  -56    & 1.279 & $B-ds^2$  &   53 $B-ds^2 \, (^5I) ^6H$   \\
	12339.04  &  12352.2    &  -13    & 1.237 & $B-ds^2$  &   36 $B-ds^2 \, (^5I) ^6H$   \\
	15136.06  &  15159.7    &  -24    & 1.170 & $B-ds^2$  &   55 $B-ds^2 \, (^5I) ^4I$   \\
	15855.28  &  15892.2    &  -37    & 1.261 & $A-sp$    &   55 $A-sp   \, (^4I) ^6H$   \\
	16154.21  &  16087.4    &   67    & 1.183 & $B-ds^2$  &   28 $B-ds^2 \, (^5I) ^6I$   \\
	16882.28  &  16900.3    &  -18    & 1.125 & $A-sp$    &   35 $A-sp   \, (^4I) ^2K$   \\
	18651.53  &  18600.1    &   51    & 1.196 & $A-sp$    &   25 $A-sp   \, (^4I) ^4Ia$  \\
	18756.22  &  18744.3    &   12    & 1.029 & $B-ds^2$  &   56 $B-ds^2 \, (^5I) ^6L$   \\
	20074.89  &  20083.9    &   -9    & 1.094 & $B-ds^2$  &   35 $B-ds^2 \, (^5I) ^6K$   \\
	22227.34  &  22221.5    &    6    & 1.153 & $A-sp$    &   42 $A-sp   \, (^4I) ^6K$   \\
	23445.28  &  23451.8    &   -6    & 1.010 & $B-ds^2$  &   59 $B-ds^2 \, (^5I) ^4L$   \\
	23834.94  &  23842.2    &   -7    & 1.176 & $A-sp$    &   43 $A-sp   \, (^4I) ^6I$   \\
	24660.80  &  24723.4    &  -63    & 1.189 & $A-sp$    &   42 $A-sp   \, (^4I) ^4Ib$  \\
	25272.63  &  25270.0    &    3    & 1.299 & $B-ds^2$  &   76 $B-ds^2 \, (^5F) ^6H$   \\
	26957.88  &  27015.9    &  -58    & 1.118 & $A-sp$    &   25 $A-sp   \, (^4I) ^6K$   \\
	29642.95  &  29658.5    &  -16    & 1.093 & $A-sp$    &   48 $A-sp   \, (^4I) ^4Kb$  \\
	37233.47  &  37253.2    &  -20    & 1.153 & $B-ds^2$  &   16 $B-ds^2 \, (^3L) ^4I$   \\
	\hline
	& & & $J=17/2$ & &     \\
	8378.91  &   8370.5     &    8    & 1.267 & $B-ds^2$  &   73 $B-ds^2 \, (^5I) ^6I$   \\
	11530.56  &  11523.0    &    8    & 1.190 & $B-ds^2$  &   52 $B-ds^2 \, (^5I) ^4K$   \\
	15130.31  &  15152.3    &  -22    & 1.148 & $B-ds^2$  &   33 $B-ds^2 \, (^5I) ^6L$   \\
	16438.01  &  16421.3    &   17    & 1.147 & $B-ds^2$  &   36 $B-ds^2 \, (^5I) ^6K$   \\
	16709.82  &  16640.0    &   70    & 1.229 & $A-sp$    &   40 $A-sp   \, (^4I) ^6I$   \\
	18337.80  &  18305.2    &   33    & 1.239 & $A-sp$    &   52 $A-sp   \, (^4I) ^6I$   \\
	20568.63  &  20589.1    &  -21    & 1.099 & $B-ds^2$  &   47 $B-ds^2 \, (^5I) ^4L$   \\
	23498.57  &  23549.6    &  -51    & 1.197 & $A-sp$    &   60 $A-sp   \, (^4I) ^6K$   \\
	24360.81  &  24354.1    &    7    & 1.176 & $A-sp$    &   57 $A-sp   \, (^4I) ^4Ka$  \\   
	\hline
	& & & $J=19/2$ & &     \\
	9741.50  &   9772.5    &  -31    & 1.230 & $B-ds^2$  &   67 $B-ds^2 \, (^5I) ^6K$   \\
	11689.77  &  11650.3    &   40    & 1.181 & $B-ds^2$  &   57 $B-ds^2 \, (^5I) ^4L$   \\
	16683.52  &  16718.0    &  -34    & 1.176 & $B-ds^2$  &   69 $B-ds^2 \, (^5I) ^6L$   \\
	17883.57  &  17897.5    &  -14    & 1.261 & $A-sp$    &   97 $A-sp   \, (^4I) ^6K$   \\
	\hline
	& & & $J=21/2$ & &     \\
	11322.31  &  11358.3    &  -36    & 1.232 & $B-ds^2$  &   93 $B-ds^2 \, (^5I) ^6L$   \\	 
	\hline
\end{longtable*}

\begin{longtable*}{crrrrrrrrr}
	\caption{\label{tab:parev} Fitted parameters (in cm$^{-1}$) for even-parity configurations of Ho I compared	with relativistic Hartree-Fock (HFR) values. The scaling factors are  $SF(P)=P_\mathrm{fit}/P_\mathrm{HFR}$, except for $E_\mathrm{av}$ where they equal $P_{fit}-P_{HFR}$. Some parameters are constrained to vary in a constant ratio $r_n$, indicated in the second column except if {}`fix' appears in the second or in the {}`Unc.' columns. In this case, the parameter $P$ is not adjusted. The {}`Unc.' columns named after {}`uncertainty' present the standard error on each parameter after the fitting procedure.}
	\\
	\hline
	& &   \multicolumn{4}{c}{4f$^{10}5d6s^2$} & \multicolumn {4}{c}{4f$^{11}6s6p$} \\
	\cline{3-6} \cline{7-10}
	\endfirsthead
	\caption{Fitted parameters in Ho I (continued)} \\
	\hline
	& & \multicolumn{4}{c}{Fitted parameters} & \multicolumn{4}{c}{Fitted parameters} \\
	\cline{3-6} \cline{7-10}
	Param. $P$      &  Cons.   & $P_\mathrm{fit}$ & Unc. & $P_\mathrm{HFR}$   & $SF$ &  $P_\mathrm{fit}$  & Unc.   & $P_\mathrm{HFR}$   & $SF$ \\
	\hline
	\endhead                                                                  
	Param. $P$      &  Cons.   & $P_\mathrm{fit}$ & Unc. & $P_\mathrm{HFR}$ &  $SF$  &  $P_\mathrm{fit}$ & Unc.  & $P_\mathrm{HFR}$ &  $SF$ \\
	\hline
	$E_\mathrm{av}$ &          &           59617  & 105  &            5940  & 53677  &            51079 &   64  &            15134 & 35945 \\
	$F^2(4f4f)$     &  $r_{1}$ &           94927  & 707  &          125792  & 0.755  &            89432 &  666  &           118509 & 0.755 \\
	$F^4(4f4f)$     &  $r_{2}$ &           66088  & 1446 &           78881  & 0.838  &            61977 &  1356 &            73975 & 0.838 \\
	$F^6(4f4f)$     &  $r_{3}$ &           48377  & 1350 &           56738  & 0.853  &            45289 &  1264 &            53115 & 0.853 \\
	$\alpha$        &  $r_{4}$ &            23.0  &   4  &                  &        &             23.0 &    4  &                  &       \\
	$\beta$         &      fix &            -650  &      &                  &        &             -650 &       &                  &       \\
	$\gamma$        &      fix &            2000  &      &                  &        &             2000 &       &                  &       \\
	$\zeta_{4f}$    &  $r_{5}$ &            2141  &   4  &           2193   & 0.976  &             2009 &    4  &             2058 & 0.976 \\ 
	$\zeta_{5d}$    &  $r_{7}$ &             757  &  11  &            920   & 0.823  &                  &       &                 &      \\   
	$\zeta_{6p}$    & $r_{16}$ &                  &      &                  &        &             1435 &   15  &              990 & 1.449 \\ 
	$F^1(4f5d)$     &  $r_{8}$ &             674  &  91  &                  &        &                  &       &                  &       \\ 
	$F^2(4f5d)$     &  $r_{9}$ &           15491  & 279  &           20639  & 0.751  &                  &       &                  &       \\ 
	$F^4(4f5d)$     & $r_{10}$ &           10954  & 464  &            9423  & 1.162  &                  &       &                  &      \\
	$F^1(4f6p)$     &      fix &                  &      &                  &        &              150 &       &                  &     \\   
	$F^2(4f6p)$     & $r_{17}$ &                  &      &                  &        &             3643 &  289  &             3324 & 1.096  \\
	$G^1(4f5d)$     & $r_{11}$ &            5410  & 151  &            8944  & 0.605  &                  &       &                 & \\        
	$G^2(4f5d)$     & $r_{12}$ &            1378  & 434  &                  &        &                  &       &                 &  \\       
	$G^3(4f5d)$     & $r_{13}$ &            6036  & 460  &            7086  & 0.852  &                  &       &                 &  \\
	$G^4(4f5d)$     & $r_{14}$ &            2314  & 546  &                  &        &                  &       &                 & \\        
	$G^5(4f5d)$     & $r_{15}$ &            4508  & 306  &            5353  & 0.842  &                  &       &                 &      \\
	$G^3(4f6s)$     & $r_{18}$ &                  &      &                  &        &             1358 &   92  &             1676 & 0.810 \\
	$G^2(4f6p)$     &      fix &                  &      &                  &        &              760 &       &              760 & 1.0 \\        
	$G^4(4f6p)$     &      fix &                  &      &                  &        &              662 &       &             662 & 1.0  \\ 
	$G^1(6s6p)$     & $r_{19}$ &                  &      &                  &        &            10321 &   74  &           23282 & 0.443 \\
	configuration-interaction & & \multicolumn{4}{c}{$4f^{10}5d6s^2-4f^{11}6s6p$} \\
	\hline
	\cline{3-6}
	$R^{1}(5d6s,4f6p)$ & $r_{6}$ &  -3223 &  150 &  -4555 & 0.708  \\
	$R^{3}(5d6s,6p4f)$ & $r_{6}$ &   -685 &   32 &   -968 & 0.708  \\                                                        
	\hline		
\end{longtable*}

\section*{Bibliography}


%

\end{document}